\documentclass[superscriptaddress,showpacs,amssymb,10pt,reprint,aps,prd,longbibliography,nofootinbib,floatfix]{revtex4-1}

\usepackage{graphicx,epsfig,amssymb,times} 
\usepackage{amsmath,amsfonts}
\usepackage{bm}
\usepackage{epstopdf}
\usepackage{hyperref}
\usepackage[caption=false]{subfig}
\usepackage[usenames]{color}   
\usepackage[dvipsnames]{xcolor}

\usepackage[normalem]{ulem}

\definecolor{coolblack}{rgb}{0.0, 0.18, 0.39}
\definecolor{darkred}{rgb}{0.5,0,0}
\definecolor{darkgreen}{rgb}{0,0.5,0}
\definecolor{darkblue}{rgb}{0,0,0.5}
\definecolor{lapislazuli}{rgb}{0.15, 0.38, 0.61}
\definecolor{venetianred}{rgb}{0.78, 0.03, 0.08}
\definecolor{bleudefrance}{rgb}{0.19, 0.55, 0.91}
\definecolor{dogwoodrose}{rgb}{0.84, 0.09, 0.41}
\hypersetup{colorlinks=true, citecolor=darkblue, linkcolor=darkblue, 
urlcolor = darkblue}

\def\be{\begin{equation}}
\def\ee{\end{equation}}

\newcommand{\bea}{\begin{eqnarray}}
\newcommand{\eea}{\end{eqnarray}}
\newcommand{\ben}{\begin{enumerate}}
\newcommand{\een}{\end{enumerate}}
\newcommand{\bi}{\begin{itemize}}
\newcommand{\ei}{\end{itemize}}

\def\ga{\mathrel{\raise.3ex\hbox{$>$\kern-.75em\lower1ex\hbox{$\sim$}}}}
\def\la{\mathrel{\raise.3ex\hbox{$<$\kern-.75em\lower1ex\hbox{$\sim$}}}}

\def\be{\begin{equation}}
\def\ee{\end{equation}}

\def\I_M{{I_{\scriptscriptstyle M\times M}}}

\def\be{\begin{equation}}
\def\ee{\end{equation}}
\def\bea{\begin{eqnarray}}
\def\eea{\end{eqnarray}}
\newcommand{\beq}{\begin{eqnarray}}
\newcommand{\eeq}{\end{eqnarray}}

\newcommand{\beqal}{\begin{eqnarray}\label}
\newcommand{\beqa}{\begin{eqnarray}}
\newcommand{\eeqa}{\end{eqnarray}}

\begin{document}
\title{\large Electrically charged black holes in linear and nonlinear electrodynamics: Geodesic analysis and scalar absorption}
	
	\author{Marco A. A. Paula}
	\email{marco.paula@icen.ufpa.br}
	\affiliation{Programa de P\'os-Gradua\c{c}\~{a}o em F\'{\i}sica, Universidade 
		Federal do Par\'a, 66075-110, Bel\'em, Par\'a, Brazil.}

\author{Luiz C. S. Leite}
\email{luiz.leite@ifpa.edu.br}
\affiliation{Programa de P\'os-Gradua\c{c}\~{a}o em F\'{\i}sica, Universidade 
	Federal do Par\'a, 66075-110, Bel\'em, Par\'a, Brazil.}
\affiliation{Campus Altamira, Instituto Federal do Par\'a, 68377-630, Altamira, Par\'a, Brazil.}
	
\author{Lu\'is C. B. Crispino}
\email{crispino@ufpa.br}
\affiliation{Programa de P\'os-Gradua\c{c}\~{a}o em F\'{\i}sica, Universidade 
		Federal do Par\'a, 66075-110, Bel\'em, Par\'a, Brazil.}

\begin{abstract}
Along the last decades, several regular black hole (BH) solutions, i.e., singularity-free BHs, have been proposed and associated to nonlinear electrodynamics models minimally coupled to general relativity. Within this context, it is of interest to study how those nonlinear-electrodynamic-based regular BHs (RBHs) would interact with their astrophysical environment. We investigate the propagation of a massless test scalar field in the background of an electrically charged RBH solution, obtained by Eloy Ay\'on-Beato and Alberto Garc\'ia. Using a numerical approach, we compute the absorption cross section of the massless scalar field for arbitrary values of the frequency of the incident wave. We compare the absorption cross sections of the Ay\'on-Beato and Garc\'ia RBH with the Reissner-Nordstr\"om BH, showing that they can be very similar in the whole frequency regime.
\end{abstract}

\date{\today}

\maketitle

\section{Introduction}
General relativity (GR) is a very well-established gravitational theory that has successfully passed through many experimental tests~\cite{W2014} and also predicted new astrophysical objects and phenomena, like black holes (BHs)~\cite{AAA2019L1} and gravitational waves~\cite{AAA2016}. Within GR, standard BHs are characterized by an event horizon and described by only three parameters, namely the following: the mass, the electric charge and the angular momentum~\cite{H1996}. However, despite this simplicity, the curvature singularities, hidden inside the BH event horizon according to the cosmic censorship conjecture~\cite{W1984}, represent a potential challenge to GR.

During the last 50 years, many efforts have been made to circumvent the problem of intrinsic singularities within GR, including the so-called regular BH (RBH) solutions (for a review see, e.g., Ref.~\cite{A2008}). Historically, the first suggested RBHs lacked of a specified source associated to their line elements (see, e.g., Refs.~\cite{B1968,B1994,BF1996,MMPS1996,CAB1997} and references therein). However, in 1998, Eloy Ay\'on-Beato and Alberto Garc\'ia proposed a nonlinear electrodynamics (NED) model minimally coupled to GR~\cite{ABG1998} as a possible source to singularity-free charged BHs. The NED generalizes Maxwell's theory~\cite{B1934,BI1934,P1970} and appears at certain energy levels of some string/M theories~\cite{FT1985,SW1999,A2000}. Based on a NED framework, several electrically~\cite{ABG1999,ABG1999-2,D2004,BV2014,RS2018} and magnetically~\cite{ABG2000,B2001,M2004,M2015,K2017} charged RBH solutions have been proposed, as well as NED-based RBH solutions in alternative theories of gravity~\cite{JRH2015,SR2018}.

It is well known that, in real astrophysical scenarios, BHs are surrounded by distributions of matter~\cite{N2005}. Within this context, in order to improve our understanding on BH physics, we can study how BHs absorb and scatter matter fields. Many investigations concerning the absorption and scattering have been made for standard BH solutions (see, e.g., Refs.~\cite{FHM1988,OCH2011,CDHO2014,CDHO2015,BC2016,LBC2017} and references therein). Recently, some studies related to how test matter fields are absorbed and scattered by RBHs (in the NED framework) have also been carried out~\cite{MC2014,MOC2015,SBP2017,S2017}, but some features are yet to be investigated. For instance, in the scenarios of test scalar fields absorption, the role played by the RBH's electric charge and the possibility of such electrically charged RBHs mimic the standard BHs.

We study the absorption of a massless test scalar field in the background of the RBH solution obtained by Ay\'on-Beato and Garc\'ia (ABG)~\cite{ABG1998}, which is a static, spherically symmetric, and electrically charged RBH. By using a numerical approach we compute the absorption cross section (ACS) for arbitrary values of the field frequency, and we also perform a classical analysis of the ACS. Noting that the ABG RBH has a causal structure similar to that of the Reissner-Nordstr\"om (RN) BH, we compare our results with the RN ones~\cite{JP2005,CDE2009}.

The remainder of this paper is organized as follows. In Sec.~\ref{sec:RBHspacetime} we review the main aspects of the ABG RBH spacetime. We perform a classical analysis of the absorption of massless particles in Sec.~\ref{sec:ca}, and in Sec~\ref{sec:sf} we study the dynamics of a massless scalar field in the background of the ABG RBH. In Sec.~\ref{sec:acs} we investigate the ACS using the partial-wave method and exhibit approximations for the low- and high-frequency regime. In Sec.~\ref{sec:results} we present our main results associated to the ACS of the ABG RBH. We conclude with our final remarks in Sec.~\ref{sec:remarks}. Throughout this paper we use natural units, for which $G = c = \hbar = 1$, and the metric signature $(+,-,-,-)$.

\section{ABG RBH spacetime}\label{sec:RBHspacetime}
The NED theory (in the so-called $P$ framework~\cite{B2001}) minimally coupled to GR can be described by the action
\begin{equation}
\label{S}\mathrm{S} = \dfrac{1}{4 \pi}\int d^{4}x \sqrt{-g} \left[ \dfrac{1}{4}R-\left(2P\mathcal{H}_{P}-\mathcal{H}(P)\right) \right], 
\end{equation}
where $g$ is the determinant of the metric tensor $g^{\mu \nu}$, $R$ is the corresponding Ricci scalar, $\mathcal{H}(P)$ is a Hamiltonian-like density quantity obtained through a Legendre transformation~\cite{HGP1987}, and $\mathcal{H}_{P} \equiv \partial\mathcal{H}/\partial P$. The auxiliary antisymmetric tensor $P_{\mu \nu}$ and the scalar $P$ are given by
\begin{equation}
\label{EFT}P_{\mu \nu} = \mathcal{H}_{P}^{-1}F_{\mu \nu} \ \ \ \text{and} \ \ \ P \equiv \dfrac{1}{4}P_{\mu \nu}P^{\mu \nu},
\end{equation}
respectively, with $F_{\mu\nu}$ being the standard electromagnetic field strength. A correspondence with NED theory in the $F$ framework can be obtained considering the following relations~\cite{HGP1987}:
\begin{equation}
\label{FPDUALITY}\mathcal{L} = 2P\mathcal{H}_{P}-\mathcal{H}(P) \, , \ \ \mathcal{L}_{F}\mathcal{H}_{P} = 1 \, , \ \ \text{and} \ \ F = P \mathcal{H}_{P}^{2} \,\, ;
\end{equation}
in which $\mathcal{L}$ is a gauge-invariant electromagnetic Lagrangian density and $\mathcal{L}_{F} \equiv \partial \mathcal{L}/\partial F$, where $F$ is the Maxwell scalar,
\begin{equation}
\label{MS}F = \dfrac{1}{4}F_{\mu\nu}F^{\mu\nu}.
\end{equation}

For the RBH solution with mass $M$ and electric charge $Q$ obtained in Ref.~\cite{ABG1998}, the corresponding NED source is determined by the function~\cite{ABG1998}
\begin{equation}
	\mathcal{H}(P) = P\frac{(1-3\sqrt{-2Q^2 P})}{(1+\sqrt{-2Q^2 P})^3}-\frac{3M}{Q^3}\left(\frac{\sqrt{-2Q^2 P}}{1+\sqrt{-2Q^2 P}}\right)^{5/2},
\end{equation}
where the invariant $P$ is a negative quantity.

In order to solve the Einstein-NED field equations obtained from the action~\eqref{S}, one may consider a static and spherically symmetric line element 
\begin{equation}
\label{LE} ds^{2}= f(r)dt^{2}-\frac{1}{f(r)}dr^{2}-r^{2}d\Omega^{2},
\end{equation}
where $d\Omega^{2} = d\theta^2 + \sin^2\theta\, d\varphi^2$ is the line element of a unit 2-sphere, and also assume that
\begin{equation}
 P_{\mu\nu}=(\delta^{t}_{\mu}\delta^{r}_{\nu}-\delta^{t}_{\nu}\delta^{r}_{\mu})D(r). 
\end{equation} 
It can be shown that $D(r)=Q/r^2$ and $P=-Q^2/2r^4$. Finally, one can show that the metric function $f(r)$ reads
\begin{equation}
\label{MF_ABG}f(r) = f^{\rm{ABG}}(r) \equiv 1-\frac{2Mr^{2}}{(r^{2}+Q^{2})^{3/2}}+\frac{Q^{2}r^{2}}{(r^{2}+Q^{2})^{2}}. 
\end{equation}
In this paper we shall call a line element given by Eq.~\eqref{LE} with the metric function $f^{\rm{ABG}}(r)$ as the ABG line element, which has been originally obtained in Ref.~\cite{ABG1998}~\footnote{We point out that Ay\'on-Beato and Garc\'ia also obtained other line elements in the NED context~\cite{ABG1999, ABG1999-2}.}. We note that, as $r \rightarrow \infty$, 
\begin{equation}
\label{MF_RN-ABG}\lim_{r \to \infty}f^{\rm{ABG}}(r) \to f^{\rm{RN}}(r),
\end{equation}
with 
\begin{equation}
\label{MF_RN}f^{\rm{RN}}(r) \equiv 1-\dfrac{2M}{r}+\dfrac{Q_{\rm{RN}}^{2}}{r^{2}}
\end{equation}
being the metric function of the RN spacetime.

Depending on the value of the ratio~$Q/M$, the ABG RBH may possess up to two horizons, and their locations are given by 
\begin{equation}
\label{CEHs}r_{\pm} \equiv |Q|\sqrt{\left(\pm \frac{\sqrt{z(s)}}{2\sqrt{6}s}+\frac{\sqrt{6u(s)+9}}{12s}+\frac{1}{4s}\right)^{2}-1},
\end{equation}
where we have used the auxiliary functions
\begin{equation}
\label{AF1}z(s) \equiv -\frac{9\left(12s^{2}-1\right)}{\sqrt{6u(s)+9}}-u(s) - 12s^{2} +3,
\end{equation}
\begin{equation}
\label{AF2}u(s) \equiv -\frac{4\left(11s^{2}-3\right)s}{\sqrt[3]{4(w(s)+9s)}}+s\sqrt[3]{4(w(s)+9s)}-4s^{2},
\end{equation}
and
\begin{equation}
\label{AF3}w(s) \equiv 74s^{3}+3\sqrt{3}\sqrt{400s^{6}-112s^{4}+47s^{2}-4},
\end{equation}
in which $s\equiv|Q|/2M$. Here we shall restrict our analysis to BHs, which are described by the ABG line element if the condition $|Q| \le Q_{\rm{ext}}\approx0.6341M$ is fulfilled~\cite{ABG1998}. When $|Q|<Q_{\rm{ext}}$ the ABG RBH possesses a Cauchy horizon at $r_{-}$ and an event horizon at $r_{+}$, given by Eq.~\eqref{CEHs}. For $|Q| = Q_{\rm{ext}}$ we have the so-called extreme ABG RBH, with $r_{+} = r_{-}$, and $|Q| > Q_{\rm{ext}}$ leads to horizonless solutions. This causal structure is similar to the RN case, for which $Q^{\rm{RN}}_{\rm{ext}}=M$.

\begin{figure}[!htbp]
\begin{centering}
    \includegraphics[width=1.0\columnwidth]{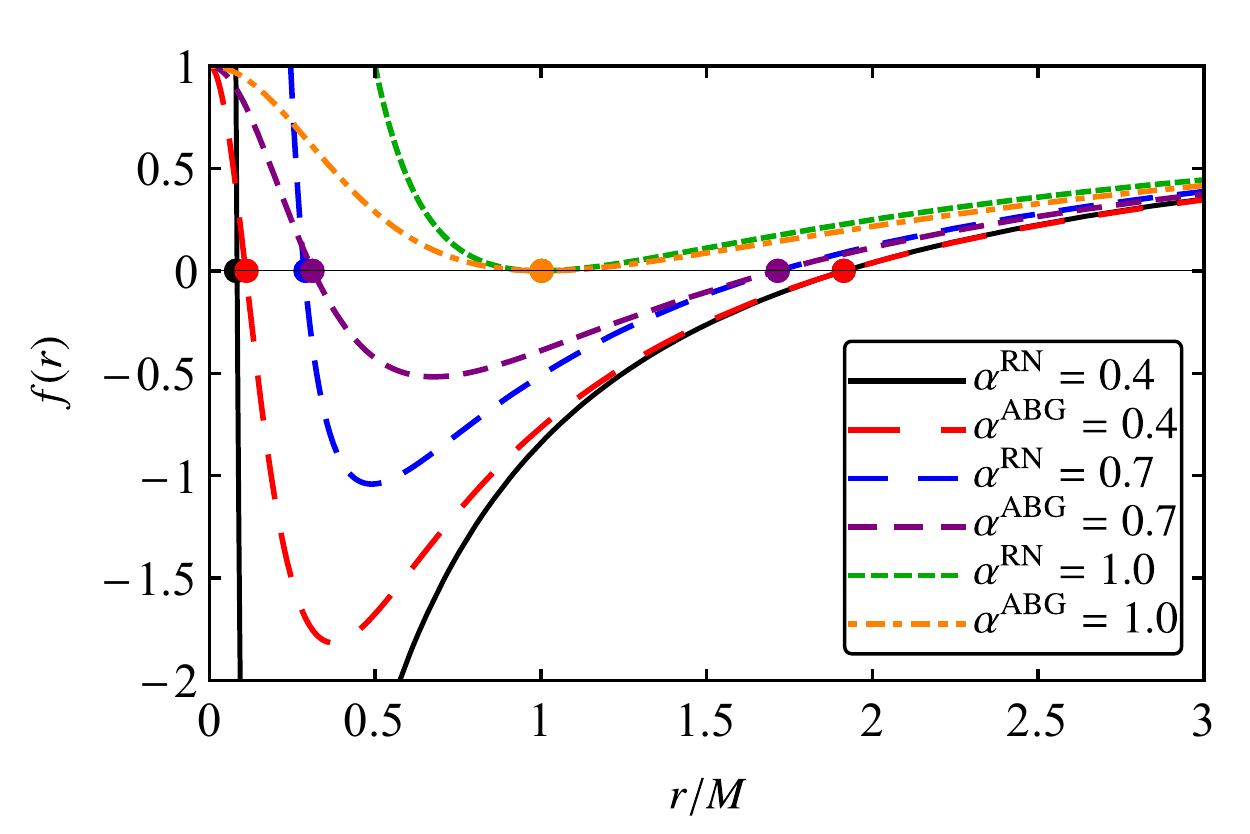}
    \caption{Comparison between the metric functions of ABG and RN BHs, as a function of $r$, for different choices of the normalized charge. The colored circles represent the location of the horizons in each case.}
    \label{comparasionfrandfrs}
\end{centering}
\end{figure}

\begin{figure}[!htbp]
\begin{center}
	\includegraphics[width=1.0\columnwidth]{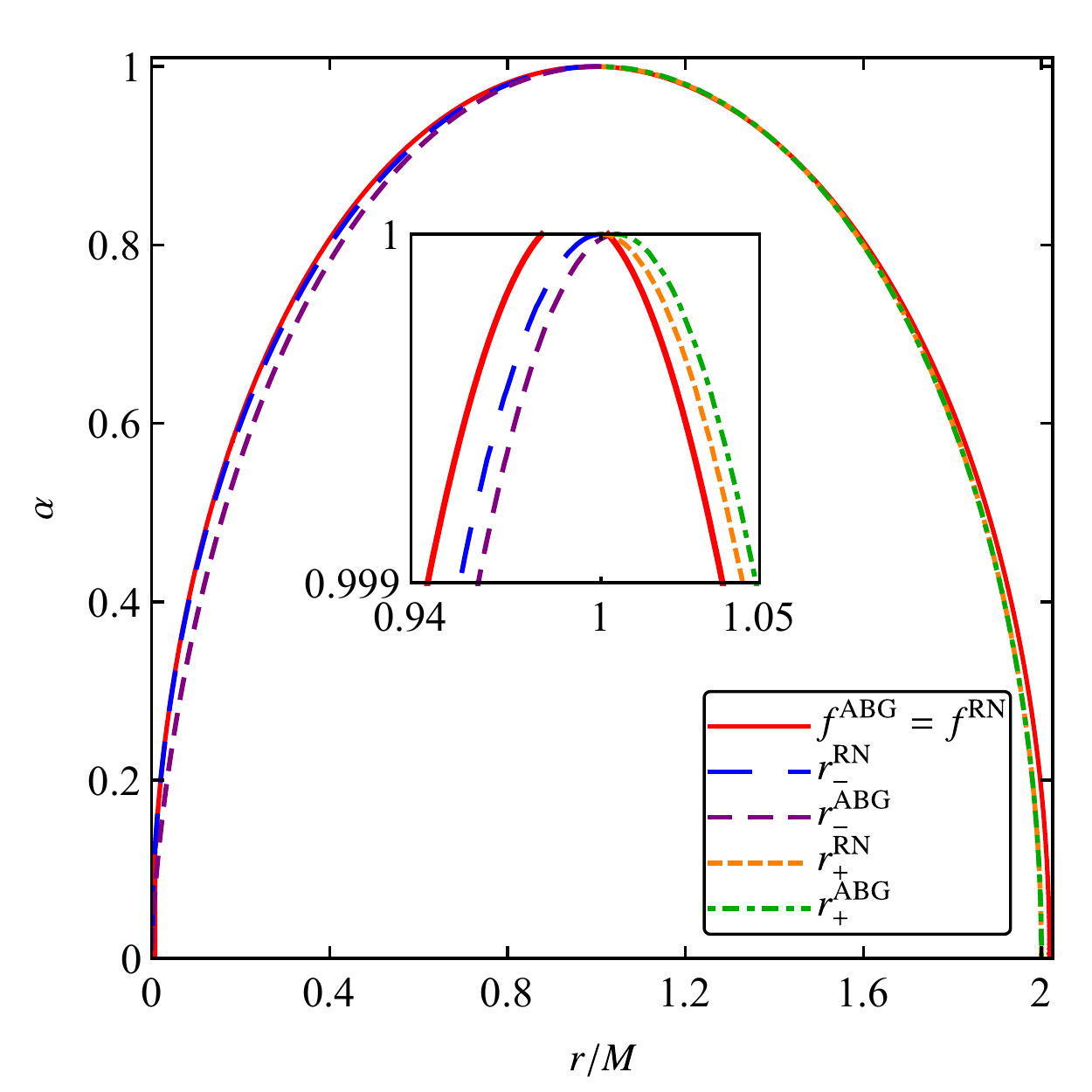}
	\caption{The horizons of ABG and RN BHs for distinct choices of the normalized charge. We also plot the function $f^{\rm{ABG}}(r) = f^{\rm{RN}}(r)$. The central inset shows the plots near the extreme charge value. Notice that there is a discontinuity at $\alpha = 1$, for the plot $f^{\rm{ABG}}(r) = f^{\rm{RN}}(r)$ .}
	\label{frnequalsfrs}
\end{center}
\end{figure}

In Fig.~\ref{comparasionfrandfrs} we plot $f(r)$, both for ABG RBHs and for RN BHs, as a function of $r$, for different values of the normalized electric charge $\alpha$, defined as 
\begin{equation}
\label{alpha}
\alpha \equiv \dfrac{Q}{Q_{\rm{ext}}}. 
\end{equation}
For fixed values of $\alpha$, we note that  $f^{\rm{ABG}}(r)$ $\rightarrow$ $f^{\rm{RN}}(r)$ as $r \rightarrow \infty$, in accordance with Eq.~\eqref{MF_RN-ABG}. 
In Fig.~\ref{frnequalsfrs} we plot the locations in which $f^{\rm{ABG}}(r) = f^{\rm{RN}}(r)$, as well as the horizons $r_+$ and $r_-$ of ABG and RN BHs, for $0 \le \alpha \le 1$. 
We note that the horizons locations for both ABG and RN BH solutions, with a fixed $\alpha$, are very similar. We also observe that the metric functions $f(r)$ of ABG and RN BHs may coincide at specific values of the radial coordinate $r$.

\begin{figure}[!htbp]
\begin{centering}
    \includegraphics[width=1.0\columnwidth]{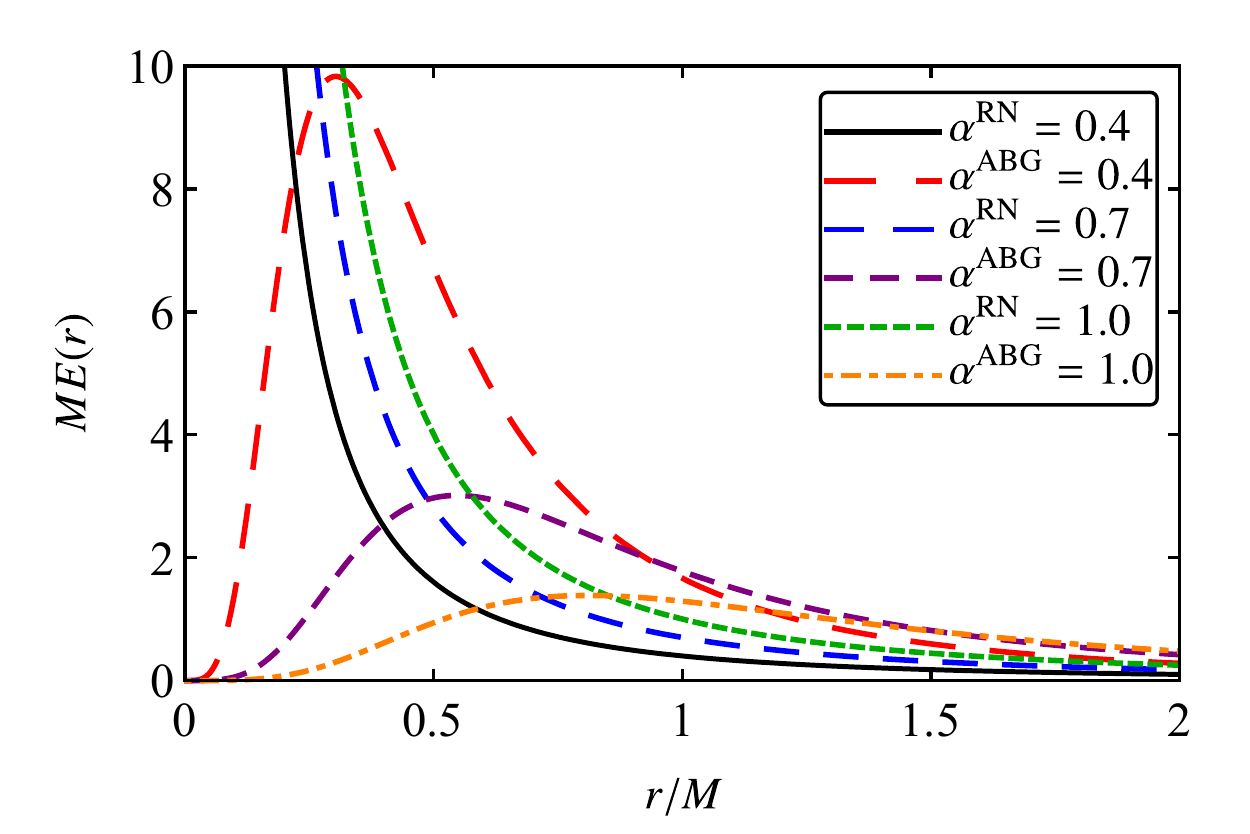}
    \caption{The electrostatic fields of ABG and RN BHs, as a function of $r$, for different values of the normalized charge.}
    \label{electricfield}
\end{centering}
\end{figure}

The (radial) electrostatic field $E(r)$ associated to the ABG solution is given by~\cite{ABG1998}
\begin{equation}
\label{E_ABG}
E^{\rm{ABG}}(r) = Qr^{4}\left(\dfrac{r^{2}-5Q^{2}}{(r^{2}+Q^{2})^{4}}+\dfrac{15M}{2(r^{2}+Q^{2})^{7/2}}\right), 
\end{equation}
which is finite at the origin (vanishing at $r = 0$) and behaves asymptotically as the electrostatic field in the RN case, namely
\begin{equation}
\label{E_RN}
E^{\rm{RN}}(r) = \dfrac{Q_{\rm{RN}}}{r^{2}}, 
\end{equation}
as it is shown in Fig.~\ref{electricfield}. We also note that, near the BH center, $E^{\rm{ABG}}(r)$ decreases as we increase $\alpha$, while the opposite behavior is observed for $E^{\rm{RN}}(r)$.

\section{Geodesic Analysis}\label{sec:ca}
In this section we obtain the classical capture cross section, also known as the geometric cross section (GCS) of null geodesics. The classical (geometric) Lagrangian related to the propagation of massless particles in the background of the line element~\eqref{LE} is given by
\begin{equation}
\label{L_SLE}2\mathrm{L}_{\rm{geo}} = g_{\mu\nu}\dot{x}^{\mu}\dot{x}^{\nu},
\end{equation}
where the overdot denotes the derivative with respect to an affine parameter $\lambda$. 
Here, due to the spherical symmetry, we can consider, without loss of generality, the motion in the equatorial plane, i.e., $\theta = \pi/2$. We can then write
\begin{equation}
\mathrm{L}_{\rm{geo}} = \frac{1}{2}\left(f(r)\dot{t}^2-\frac{\dot{r}^2}{f(r)}-r^2\dot{\varphi}^2
\right).\label{eq:lagrangian}
\end{equation}

From Eq.~\eqref{eq:lagrangian}, we note the existence of the following conserved quantities:
\begin{eqnarray}
\label{ENG}\dfrac{\partial\mathrm{L}_{\rm{geo}}}{\partial\dot{t}} = & \ f(r)\dot{t} = E,\\
\label{ANG}\dfrac{\partial\mathrm{L}_{\rm{geo}}}{\partial\dot{\varphi}} = & - r^{2}\dot{\varphi} = -L,
\end{eqnarray}
where $E$ and $L$ are the energy and the angular momentum of a massless particle, respectively.

For null geodesics, the condition $\mathrm{L}_{\rm{geo}} = 0$ has to be satisfied. Using this condition together with Eqs.~\eqref{ENG} and~\eqref{ANG}, and defining the impact parameter as
\begin{equation}
\label{IP}b \equiv \dfrac{L}{E},
\end{equation}
it is possible to obtain the following equation of motion: 
\begin{equation}
\label{CLASSICALRE} \dfrac{\dot{r}^{2}}{L^{2}} = \mathrm{h}(r) \equiv \dfrac{1}{b^{2}} - \dfrac{f(r)}{r^{2}}.
\end{equation}

From the conditions $\left.\mathrm{h}(r)\right|_{r=r_c} = 0$ and $\left.\frac{d\mathrm{h}(r)}{dr}\right|_{r=r_c} = 0$, we get the following pair of equations
\begin{align}
\label{CR}2f(r_{c})-&r_{c}f'(r_{c}) = 0,\\
\label{CIP}b_{c} = \dfrac{L_{c}}{E_{c}} &= \dfrac{r_{c}}{\sqrt{f(r_{c})}},
\end{align}
where the prime symbol, $'$, denotes the derivative with respect to the radial coordinate $r$. Using Eqs.~\eqref{CR} and~\eqref{CIP}, it is possible to find the critical radius, $r_{c}$, and the critical impact parameter, $b_{c}$, that is, the radius of an unstable circular orbit and the value for the ratio $L/E$ in the corresponding circular orbit, respectively. By solving Eq.~\eqref{CR} numerically we can compute $r_{c}$ for a given $\alpha$ and consequently find $b_{c}$. Since the GCS of null geodesics is given by $\sigma_{\rm{geo}} = \pi b_{c}^{2}$ \cite{W1984}, we obtain
\begin{equation}
\label{CACS}\sigma_{\rm{geo}} = \pi \dfrac{r_{c}^{2}}{f(r_{c})}.
\end{equation}

\begin{figure}[!htbp]
\begin{center}
	\includegraphics[width=1.0\columnwidth]{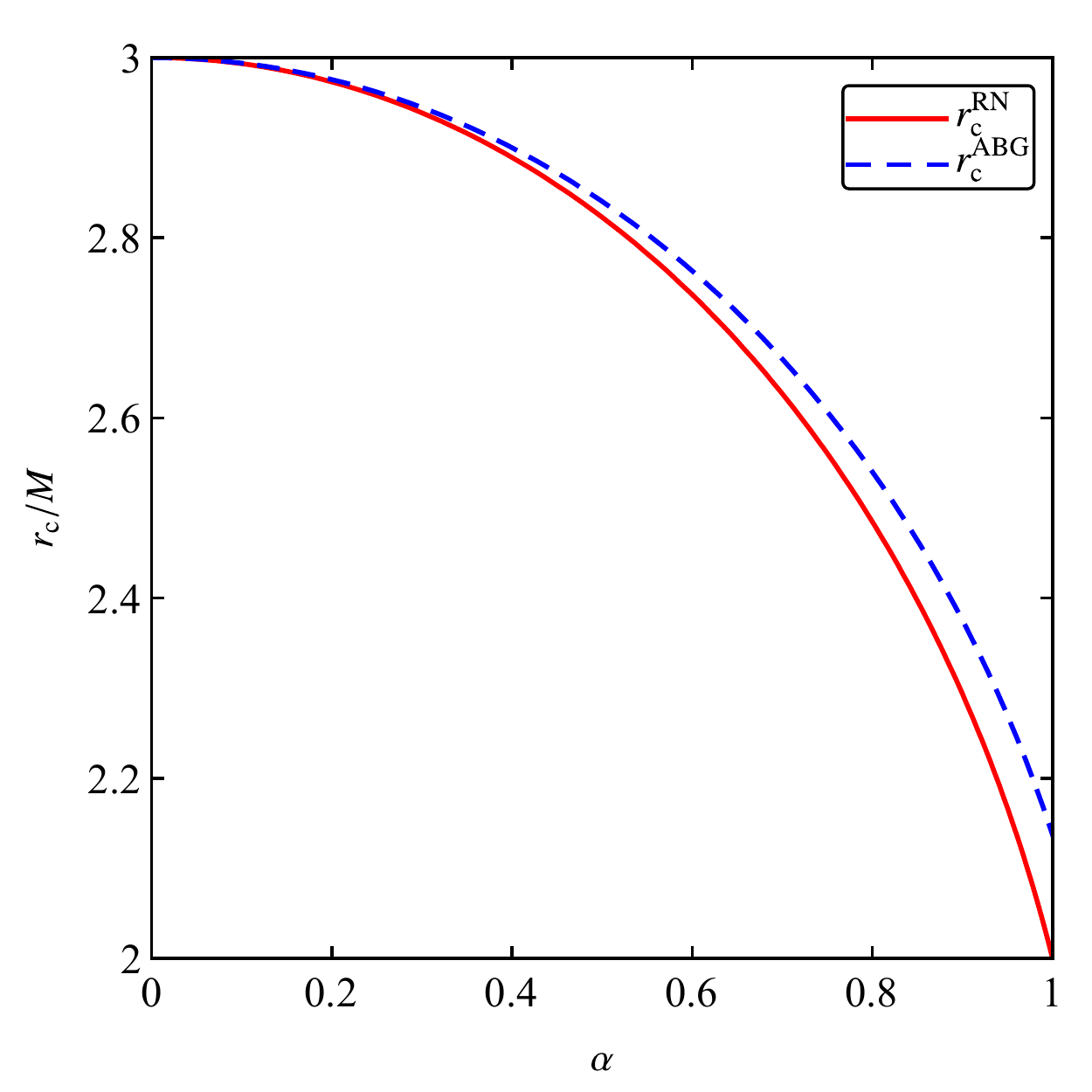}
	\includegraphics[width=1.0\columnwidth]{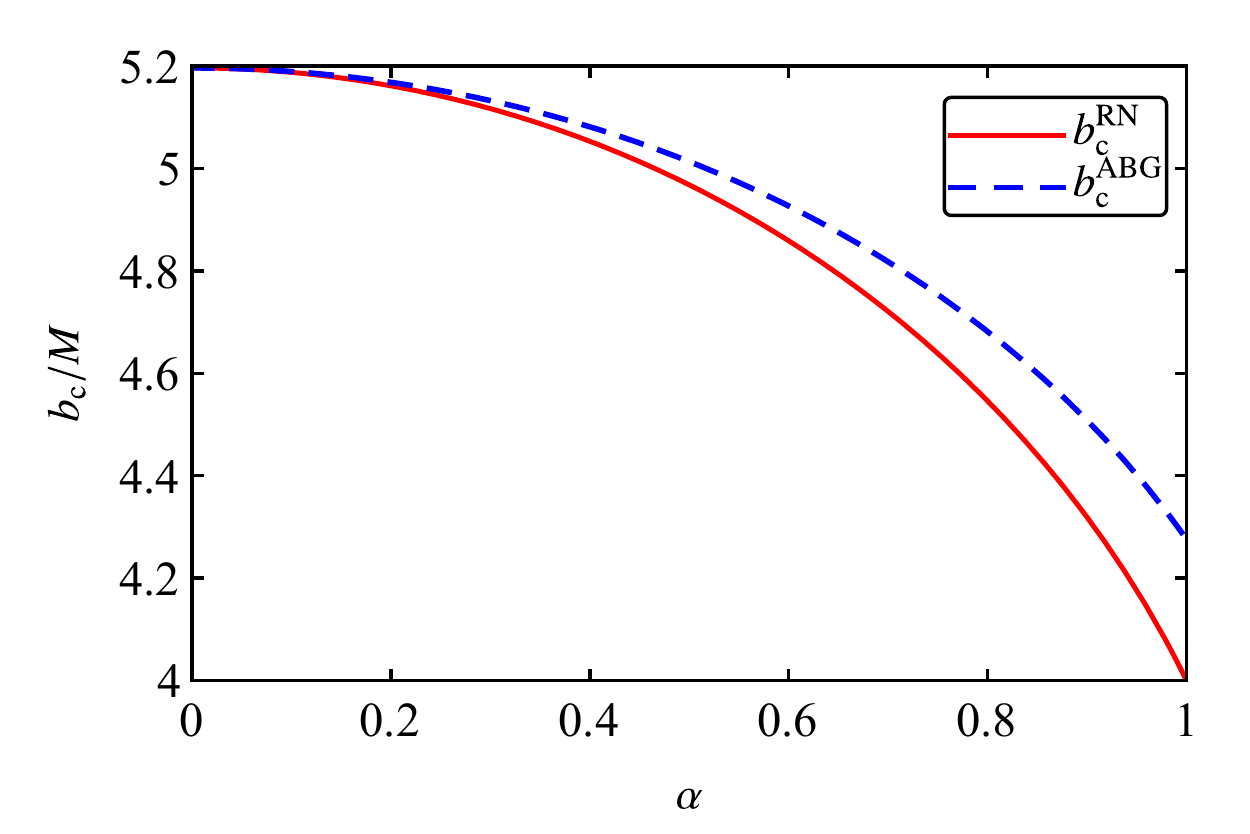}
	\caption{The critical radius (top panel) and the critical impact parameter (bottom panel) of ABG and RN BHs as functions of the normalized charge.}
	\label{criticalradius}
\end{center}
\end{figure}

In Fig.~\ref{criticalradius} we show $r_{c}$ and $b_{c}$ for ABG and RN BHs, as functions of $\alpha$. We note that, for a fixed value of $\alpha$, $r_{c}^{\rm{ABG}} > r_{c}^{\rm{RN}}$ and $b_{c}^{\rm{ABG}} > b_{c}^{\rm{RN}}$, except in the chargeless case ($\alpha = 0$), for which both results tend to the Schwarzschild values, namely $r_{c} = 3M$ and $b_{c}= 3 \sqrt{3} M$. As a consequence of the behavior presented by the critical value of the impact parameter, $b_c$, the GCS of the ABG RBH is larger than the corresponding RN BH one.

\section{Scalar field}\label{sec:sf}
\begin{figure*}[!htbp]
\begin{centering}
    \includegraphics[width=\columnwidth]{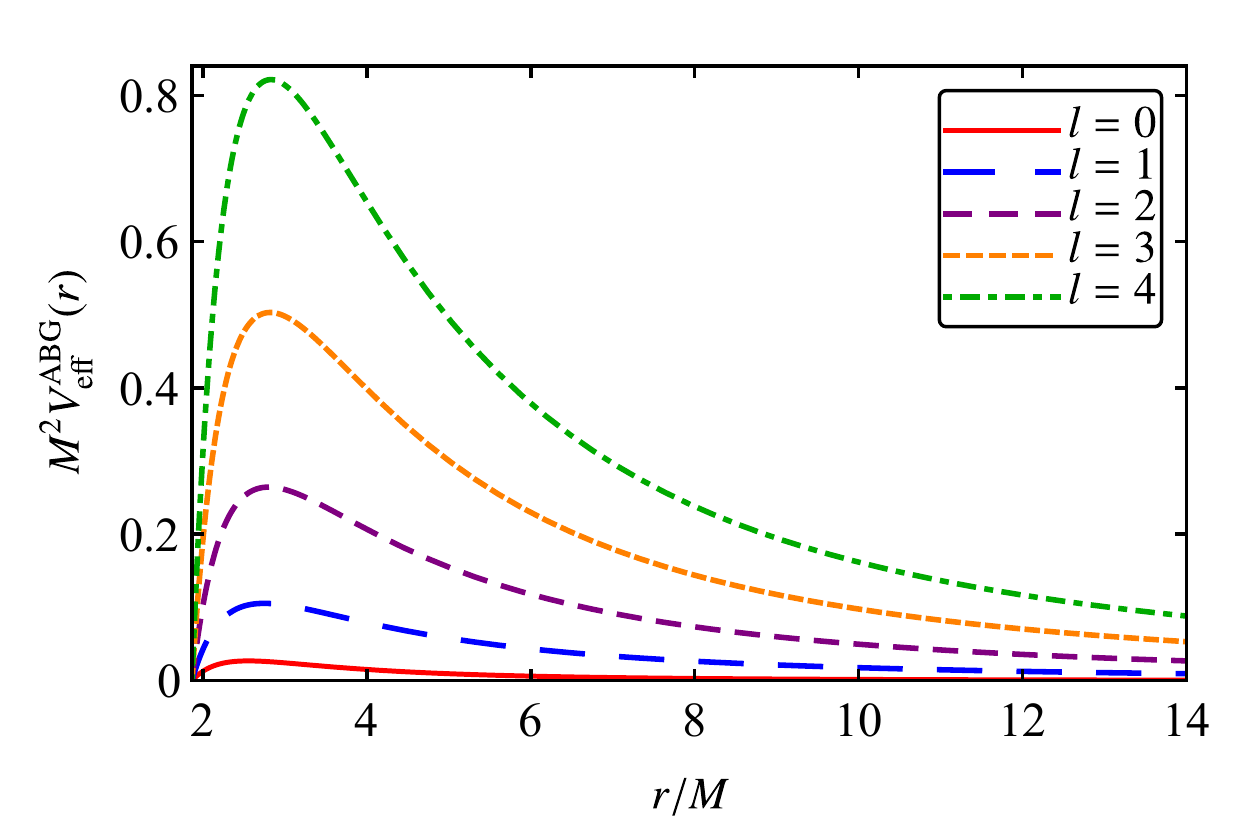}
    \includegraphics[width=\columnwidth]{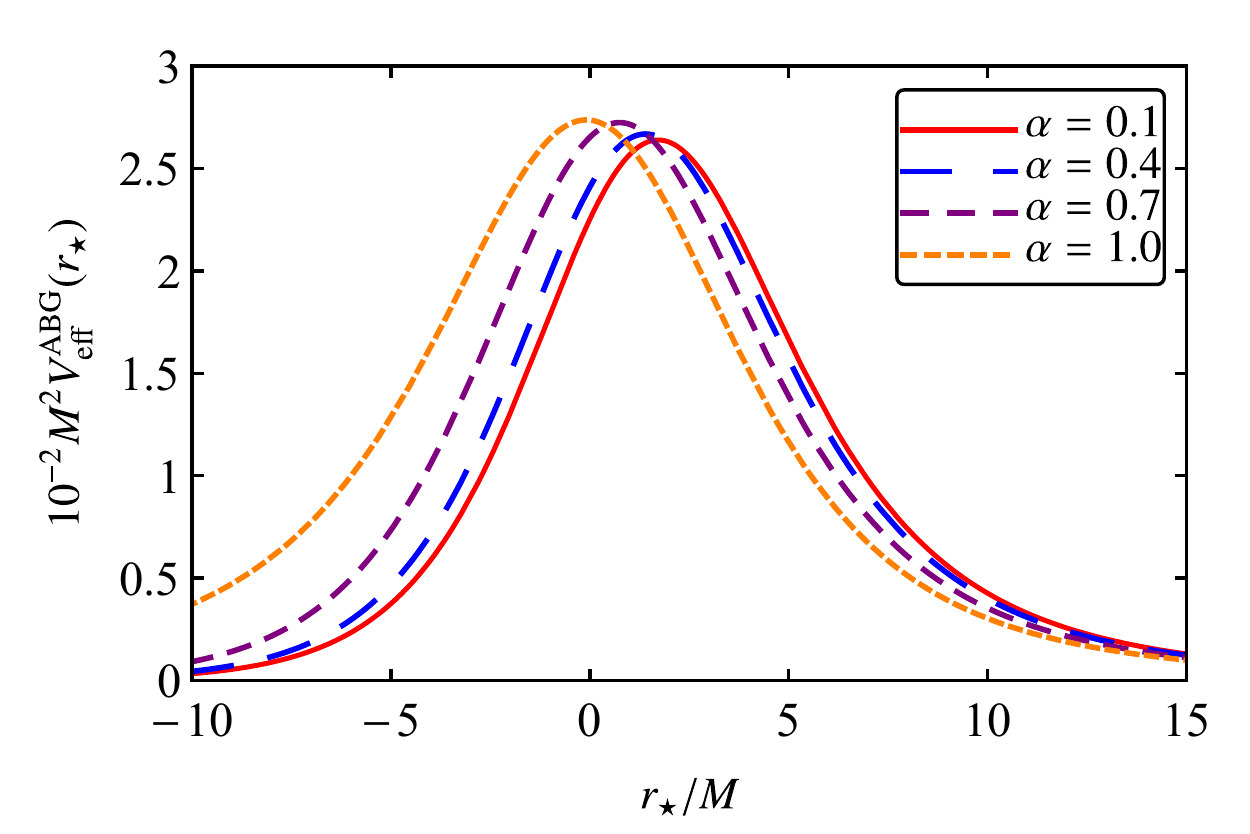}
    \caption{The effective potential of massless and chargeless scalar waves in the background of the ABG RBH: (i) as a function of $r$ for different choices of $l$, with $\alpha = 0.5$ (left panel); and (ii) as a function of $r_{\star}$ with $l = 0$, for different choices of the normalized charge (right panel).}
    \label{peff}
\end{centering}
\end{figure*}

The dynamics of a massless and chargeless test scalar field $\Phi$ is governed by the Klein-Gordon equation, i.e.,
\begin{equation}
\label{KG}\nabla_{\mu}\nabla^{\mu}\Phi = \dfrac{1}{\sqrt{-g}}\partial_{\mu}\left(\sqrt{-g}g^{\mu\nu}\partial_{\nu}\right)\Phi = 0.
\end{equation}
Considering the spherical symmetry of the spacetime under consideration, 
we can decompose $\Phi$ as
\begin{equation}
\label{PHI}\Phi\equiv \sum_{l}^{\infty} C_{\omega l}\Phi_{\omega l}=\sum_{l}^{\infty} C_{\omega l}\frac{\Psi_{\omega l}(r)}{r}P_{l}(\cos\theta)e^{-i\omega t},
\end{equation}
where $\Psi_{\omega l}(r)$ is a radial function and $P_{l}(\cos\theta)$ is the Legendre polynomial. 
The constant coefficients $C_{\omega l}$ will be determined by the boundary conditions, and the indexes $\omega$ and $l$ denote the frequency and the angular momentum of the plane wave, respectively. By inserting Eq.~\eqref{PHI} in Eq.~\eqref{KG} and defining the tortoise coordinate, $r_{\star}$, as
\begin{equation}
\label{TC}\dfrac{dr_{\star}}{dr} \equiv \dfrac{1}{f(r)},
\end{equation}
we get the following radial equation for $\Psi_{\omega l}(r)$
\begin{equation}
\label{RE} \frac{d^{2}}{dr_{\star}^{2}}\Psi_{\omega l}+\left(\omega^{2}-V_{\rm{eff}}(r)\right)\Psi_{\omega l}=0,
\end{equation}
in which the effective potential $V_{\rm{eff}}$ reads
\begin{equation}
\label{EffP}V_{\rm{eff}}(r) = f(r)\left[\dfrac{1}{r}\frac{df(r)}{dr} + \frac{l(l+1)}{r^{2}}\right].
\end{equation}
We note that the domain of the tortoise coordinate is $(-\infty,\infty)$, whereas the domain of the coordinate  $r$ is $[r_{+},\infty)$. In Fig.~\ref{peff} we present $V_{\rm{eff}}(r)$ as a function of the radial coordinate, for different choices of $l$ and $\alpha$. We note that $V_{\rm{eff}}^{\rm{ABG}}(r)$ has a peak close to $r_{+}$, which increases as we increase the values of $l$ or $\alpha$. Besides that, $V_{\rm{eff}}(r)$ vanishes in the asymptotic limits, i.e, 
\begin{equation}
\lim_{r_{\star} \rightarrow \pm \infty} V_{\rm{eff}}(r_{\star}) = 0.
\end{equation}

For the absorption/scattering problem under analysis, we consider plane waves incoming from the infinite null past (the so-called {\it in modes}). Therefore, we are interested in solutions of Eq.~\eqref{RE} subjected to the following boundary conditions 
\begin{equation}
\label{BC}\Psi_{\omega l}\sim\begin{cases}
T_{\omega l}e^{-i\omega r_{\star}}, & r\rightarrow r_{+}  \ (r_{\star}\rightarrow -\infty),\\
e^{-i\omega r_{\star}}+R_{\omega l}e^{i\omega r_{\star}}, & r\rightarrow \infty \ (r_{\star}\rightarrow \infty),
\end{cases}
\end{equation}
where $|T_{\omega l}|^2$ and $|R_{\omega l}|^2$ are the transmission and reflection coefficients, respectively. 
The plane wave subjected to the boundary conditions~\eqref{BC} comes from infinity, interacts with the effective potential~\eqref{EffP}, being partially transmitted into the BH and partially reflected back to infinity. Moreover, by using the conservation of the flux, it is possible to show that the quantities $R_{\omega l}$ and $T_{\omega l}$ satisfy
\begin{equation}
\label{CF}|R_{\omega l}|^{2}+|T_{\omega l}|^{2}=1.
\end{equation}

\section{Absorption cross section} \label{sec:acs}

\subsection{Partial-waves approach}\label{subsec:partialwaves}
In a BH absorption/scattering problem associated to static and spherically symmetric spacetimes, we consider that the field $\Phi$ behaves far from the BH as
\begin{equation}
\label{PHIACS}\Phi \sim \Phi^{\rm{(plane)}} + \Phi^{\rm{S}},
\end{equation}
where $\Phi^{\rm{(plane)}}$ is a monochromatic planar wave propagating along the $z$ axis, given by
\begin{equation}
\Phi^{\rm{(plane)}} = e^{-i\omega(t - z)},
\end{equation}
and $\Phi^{\rm{S}}$ is an outgoing scattered wave, i.e.,
\begin{equation}
\Phi^{\rm{S}} = \dfrac{1}{r}\widehat{f}(\theta)e^{-i \omega (t-r)},
\end{equation}
in which $\widehat{f}(\theta)$ is the scattering amplitude. 
We can decompose $e^{i\omega z}$ as~\cite{FHM1988}
\begin{equation}
\label{PHID}e^{i\omega z} = \sum^{\infty}_{l = 0}(2l+1)i^{l}j_{l}(\omega r)P_{l}(\cos\theta),
\end{equation}
with $j_{l}(\cdot)$ being the spherical Bessel function. 
In the far field region ($r \rightarrow \infty$), we can write
\begin{align}
\nonumber\Phi^{\rm{(plane)}} \sim \ & \dfrac{e^{- i \omega t}}{r} \sum^{\infty}_{l = 0} B_{\omega l}\left(e^{-i \omega r}+e^{-i\pi(l+1)}e^{i \omega r}\right)\times\\
\label{PHIID}& P_{l}(\cos\theta),
\end{align}
where
\begin{equation}
\label{NF}B_{\omega l} = \dfrac{(2l+1)}{2i\omega}e^{i\pi(l+1)}.
\end{equation}
If we choose a boundary condition such that the ingoing part of Eq.~\eqref{PHI} resembles, in the far field, the ingoing part of Eq.~\eqref{PHIACS}, it follows that $C_{\omega l} = B_{\omega l}$. Thus we get
\begin{equation}
\label{NEWPHI}\Phi = \sum^{\infty}_{l = 0} B_{\omega l}\Phi_{\omega l}.
\end{equation}

The ACS is related to the flux of particles transmitted into the BH. Accordingly, an expression for the ACS can be obtained by introducing the four-current density vector 
\begin{equation}
\label{CDV}J^{\mu} = \dfrac{i}{2}\left(\Phi^{*}\nabla^{\mu}\Phi-\Phi\nabla^{\mu}\Phi^{*}\right),
\end{equation}
which satisfies the conservation law $\nabla_{\mu}J^{\mu} = 0$ associated to the Klein-Gordon equation~\eqref{KG}. By inserting $\Phi_{\omega l}$ presented in Eq.~\eqref{PHI}, with $\Psi_{\omega l}$ given by Eq.~\eqref{BC} in the corresponding asymptotic limit, into Eq.~\eqref{NEWPHI} and using the orthogonality of the Legendre polynomials, i.e,
\begin{equation}
\label{OLP}\int P_{l}(\cos\theta)P_{n}(\cos\theta)d\Omega = \dfrac{4\pi}{2l+1}\delta_{ln},
\end{equation}
the surface integral of the current density vector~\eqref{CDV} leads to
\begin{equation}
\label{N}N(r) = - \int_{\Sigma} r^{2}J^{r}\text{d}\Omega = -\dfrac{\pi}{\omega}\sum_{l = 0}^{\infty}(2l+1)(1-|R_{\omega l}|^{2}),
\end{equation}
which is the flux passing thought a surface $\Sigma$ of constant radius $r$. If we consider stationary scenarios, this flux will be constant and $N$ will be (minus) the number of particles absorbed by the BH per unit of time~\cite{U1976}.

The total ACS, $\sigma(\omega)$, is defined as the ratio between the flux of $\Phi$ that goes into the BH, $|N|$, and the current of the incident planar wave, $J^{z}_{\rm{inc}} = \omega$; so that we may write
\begin{equation}
\label{TACS}\sigma(\omega) \equiv \dfrac{|N|}{J^{z}_{\rm{inc}}} = \sum_{l=0}^{\infty}\sigma_{l}(\omega),
\end{equation}
where the partial ACS, $\sigma_{l}(\omega)$, reads
\begin{equation}
\label{PACS}\sigma_{l}(\omega)=\frac{\pi}{\omega^2}(2l+1)(1-|R_{\omega l}|^{2}). 
\end{equation}

\subsection{Low- and high-frequency regimes}\label{subsec:low-frequency}
\begin{figure}[!htbp]
\begin{centering}
    \includegraphics[width=\columnwidth]{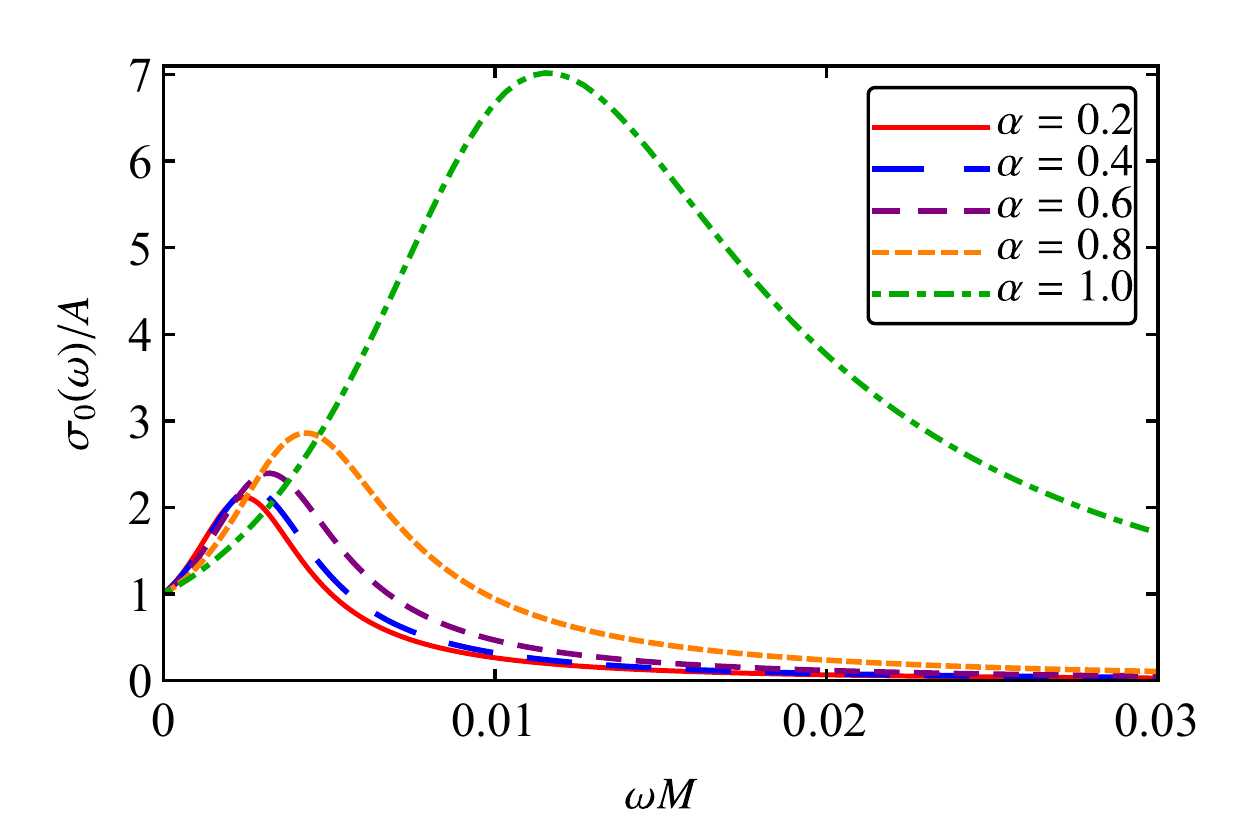}
    \caption{The ACS of the ABG RBH for a massless chargeless scalar field divided by its area, as a function of $\omega$, considering the monopole mode ($l = 0$) and different values of the normalized charge.}
    \label{lowfrequency}
\end{centering}
\end{figure}

In the low-frequency regime, it has been shown that, for stationary BH solutions, the ACS tends to the surface area of the BH event horizon~\cite{DGM1997,H2001}, which is given by
\begin{equation}
\label{BHA}A = 4\pi r_{+}^{2}. 
\end{equation}
In Fig.~\ref{lowfrequency} we show the partial ACS of the $l=0$ mode, $\sigma_{0}(\omega)$, divided by the BH area, as a function of the coupling $\omega M$. We see that, as $\omega \rightarrow 0$, the ratio $\sigma_{0}(\omega)/A$ tends to the unity, showing that, at the zero-frequency limit, the numerical result for the ACS tends to the BH area, as expected. This result can also be regarded as a consistency check of our numerical results.

\begin{figure}[!htbp]
\begin{centering}
    \includegraphics[width=\columnwidth]{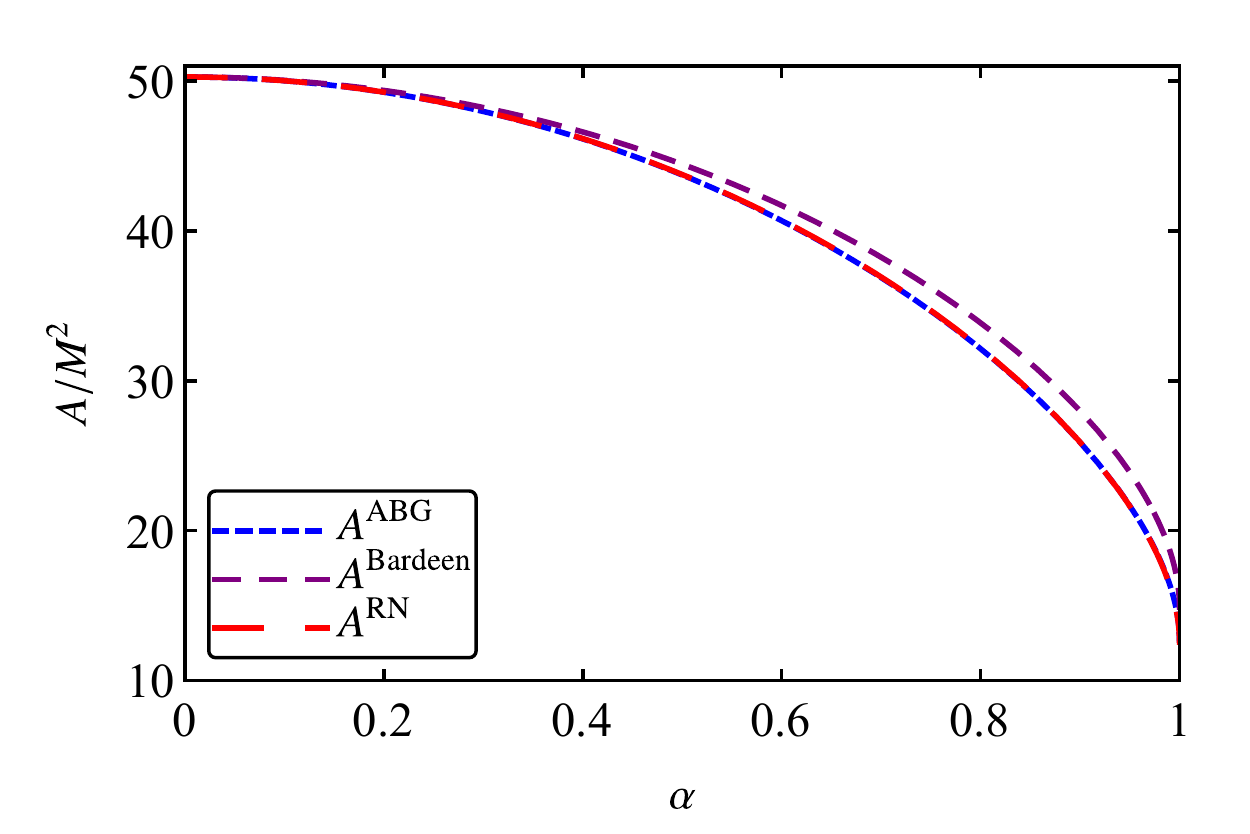}
    \caption{The horizon area of ABG, Bardeen and RN BHs as a function of the normalized charge.}
    \label{bharea}
\end{centering}
\end{figure}

In Fig.~\ref{bharea} we compare the surface area of the BH event horizon of ABG, Bardeen and RN BHs, as functions of $\alpha$. The high similarity between the value of $r_{+}$ (cf. Sec.~\ref{sec:RBHspacetime} and, in particular, Fig.~\ref{frnequalsfrs}) for ABG and for RN BHs, for the same value of $\alpha$, implies in very similar BH areas for the two cases. We also note that, for a fixed $\alpha$, the areas of ABG and RN BHs are smaller than Bardeen one.~\footnote{We take the opportunity to mention that in the caption of Fig.~10 of Ref.~\cite{MC2014} the sentence ``We have chosen $(Q_{\rm{RN}}, Q_{\rm{BD}})$ to be (0.6, 0.46809) and (0.8, 0.63252)'' should be replaced by ``We have chosen $(Q_{\rm{RN}}, Q_{\rm{BD}})$ to be (0.46809,0.6) and (0.63252,0.8)''. A similar correction is in order in the corresponding part of the text of Ref.~\cite{MC2014} in which Fig.~10 is explained.}
 
\begin{figure}[!htbp]
\begin{centering}
    \includegraphics[width=\columnwidth]{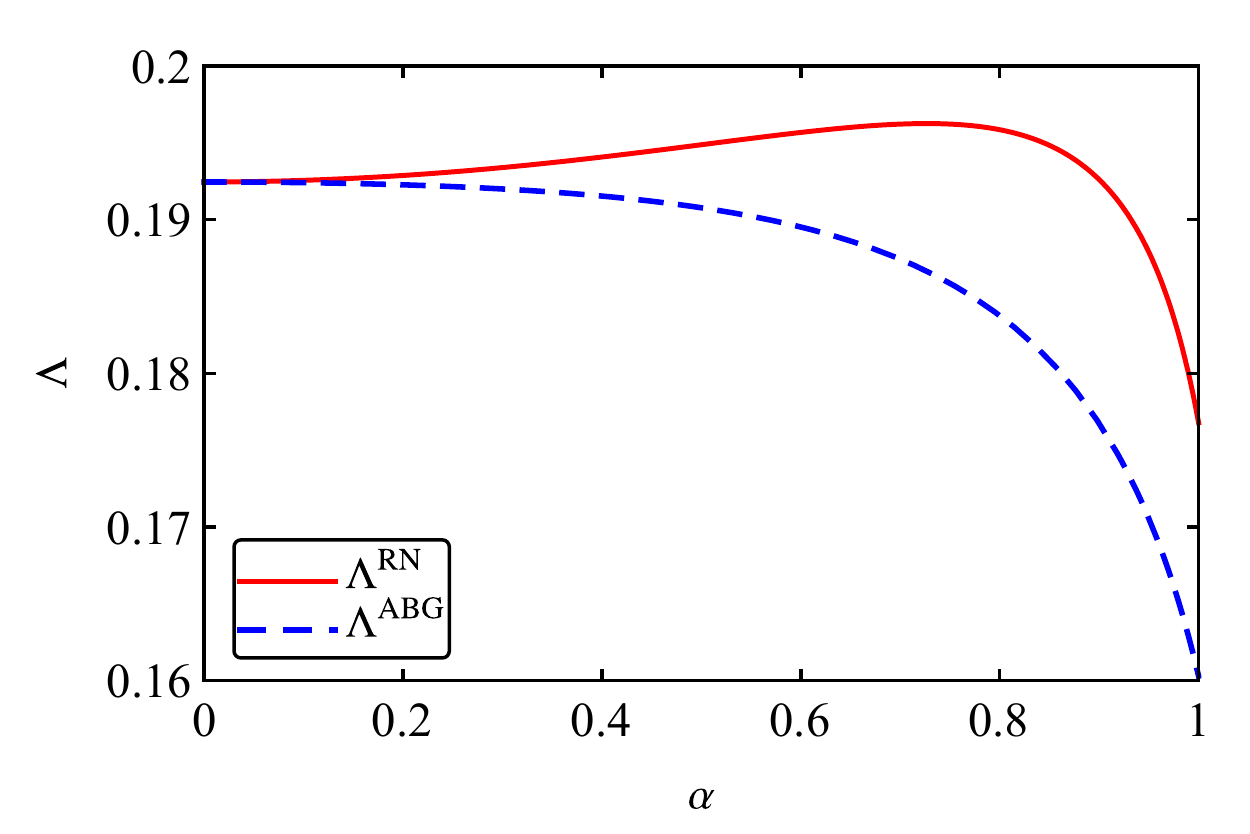}
    \caption{The Lyapunov exponent of ABG and RN BHs, as functions of the normalized charged.}
    \label{lyapunov}
\end{centering}
\end{figure}

In the high-frequency regime, massless and chargeless scalar waves can be described by null geodesics. Therefore, in this limit, the absorption of a massless scalar field is governed by Eq.~\eqref{CACS}. An improvement of the high-frequency approximation for the ACS is obtained by the so-called sinc approximation, which reveals the oscillatory behavior of the ACS. Within this approximation, the ACS can be expressed as~\cite{S1978,DEF2011}
\begin{equation}
\label{HFLIM}\sigma_{\rm{hf}} = \sigma_{\rm{geo}}\left[1-8\pi b_{c} \Lambda e^{-\pi b_{c} \Lambda}\text{sinc}\left(2\pi b_{c} \omega \right)\right] , 
\end{equation}
where $\text{sinc}(x) \equiv \sin(x)/x$, and $\Lambda$ is the Lyapunov exponent related to the unstable circular orbit~\cite{VC2009}, which is given by
\begin{equation}
\label{LEUCO}\Lambda = \sqrt{\dfrac{L^{2}_{c}}{2\dot{t}^{2}}\left(\dfrac{d^{2}\mathrm{h}(r)}{dr^{2}}\right)\bigg|_{r = r_{c}}}.
\end{equation}
The Eq.~\eqref{HFLIM} is known as the sinc approximation to the ACS. In Fig.~\ref{lyapunov} we compare Lyapunov exponent $\Lambda$ of ABG and RN BHs. We see that $\Lambda^{\rm{ABG}}$ is smaller than $\Lambda^{\rm{RN}}$, tending to the same value as $\alpha \rightarrow 0$, i.e., in the Schwarzschild BH limit. We show some results obtained using the sinc approximation in our numerical analysis in Sec.~\ref{subsec:ACSABGBH}.

\subsection{Numerical analysis}\label{subsec:numerical}
We integrate numerically Eq.~\eqref{RE} from very close to the BH event horizon $r_{+}$, up to some radial position very far from the BH, typically chosen as $r_{\infty}\sim10^{3} M$. The appropriate boundary conditions close to the $r_{+}$ and in the far field are given by Eq.~\eqref{BC}. 

With the numerical results obtained for the reflection and transmission coefficients of the scalar wave, the ACS can be computed for arbitrary values of the frequency coupling $\omega M$. For the results presented in this paper, in general, we have performed the summations in the angular momentum up to $l = 10$. The GCS and the sinc approximation are obtained using Eqs.~\eqref{CACS} and~\eqref{HFLIM}, respectively.  A selection of our numerical results is presented in Sec.~\ref{sec:results}. We have chosen to scale the ACS with the BH mass.

\section{Results}\label{sec:results}

\subsection{Absorption by the ABG RBH: main features}\label{subsec:ACSABGBH}
\begin{figure}[!htbp]
\begin{centering}
    \includegraphics[width=\columnwidth]{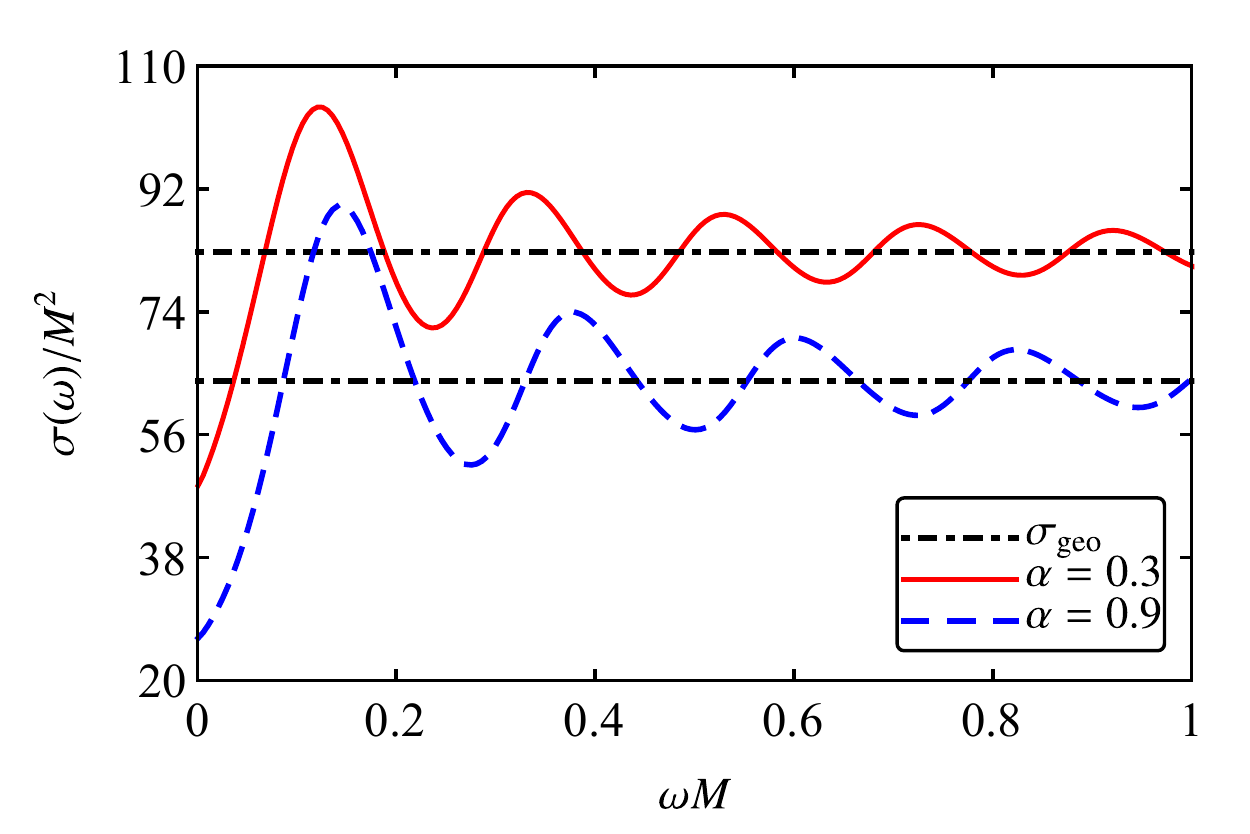}
    \includegraphics[width=\columnwidth]{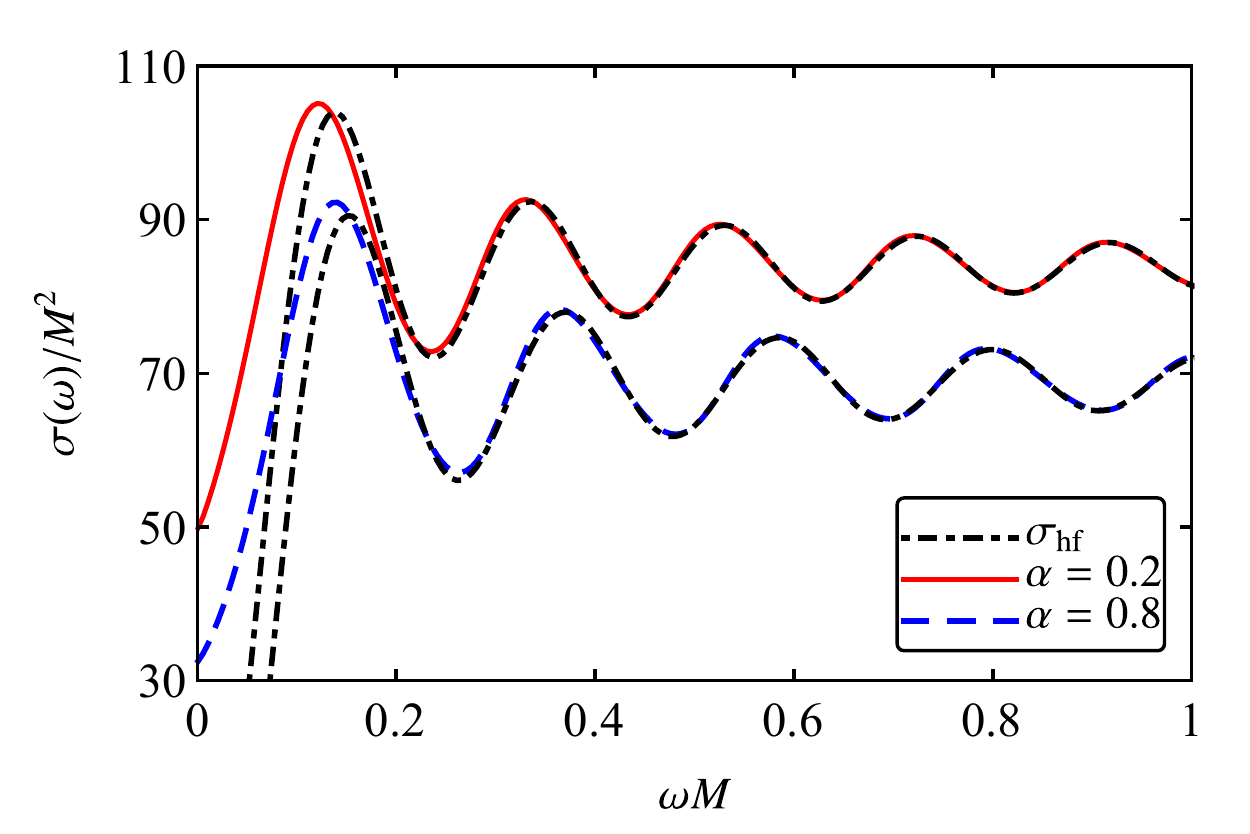}
    \caption{The total ACS of the ABG RBH for different values of the normalized charge as a function of $\omega$. In the top panel we exhibit the corresponding GCSs (horizontal dot-dashed lines), while in the bottom panel we exhibit the corresponding sinc approximation (dot-dashed lines).}
    \label{HFLIMIT}
\end{centering}
\end{figure}

\begin{figure}[!htbp]
\begin{centering}
    \includegraphics[width=\columnwidth]{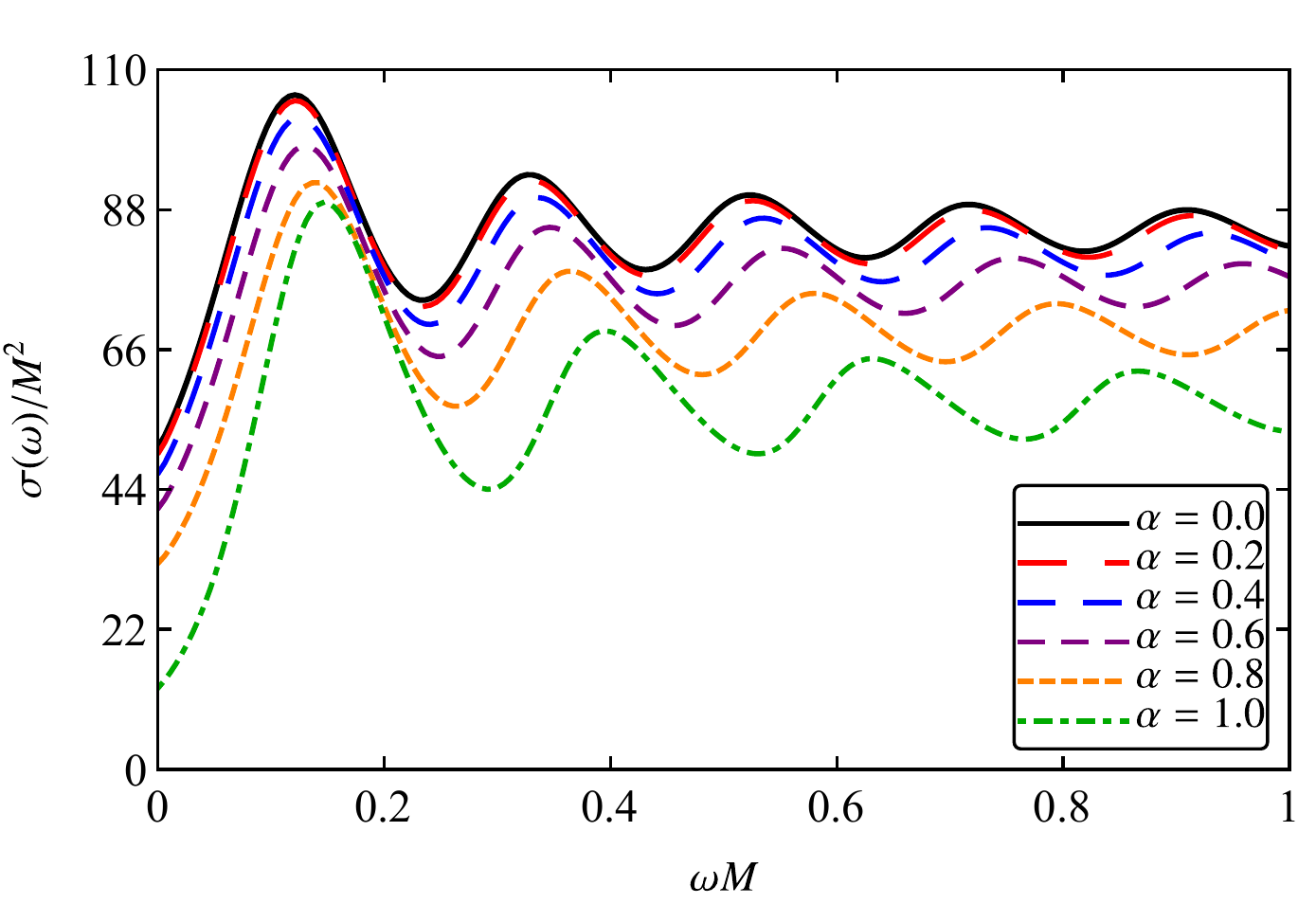}
    \caption{The total ACS of the ABG RBH for different values of normalized charge as a function of $\omega$.}
    \label{TACSQ0}
\end{centering}
\end{figure}

In Figs.~\ref{HFLIMIT} and~\ref{TACSQ0} we plot the total ACS of the ABG RBH for different values of $\alpha$ as a function of the frequency coupling $\omega M$. We note that the from mid-to-high values of the frequency the total ACS typically oscillates around the corresponding GCS (cf. top panel of Fig.~\ref{HFLIMIT}). We also observe that the sinc approximation gives an excellent approximation for the total ACS in this frequency regime (cf. bottom panel of Fig.~\ref{HFLIMIT}). We also note that the total ACS of the ABG RBH decreases as we increase $\alpha$ (cf. Fig.~\ref{TACSQ0}).

\begin{figure}[!htbp]
\begin{centering}
    \includegraphics[width=\columnwidth]{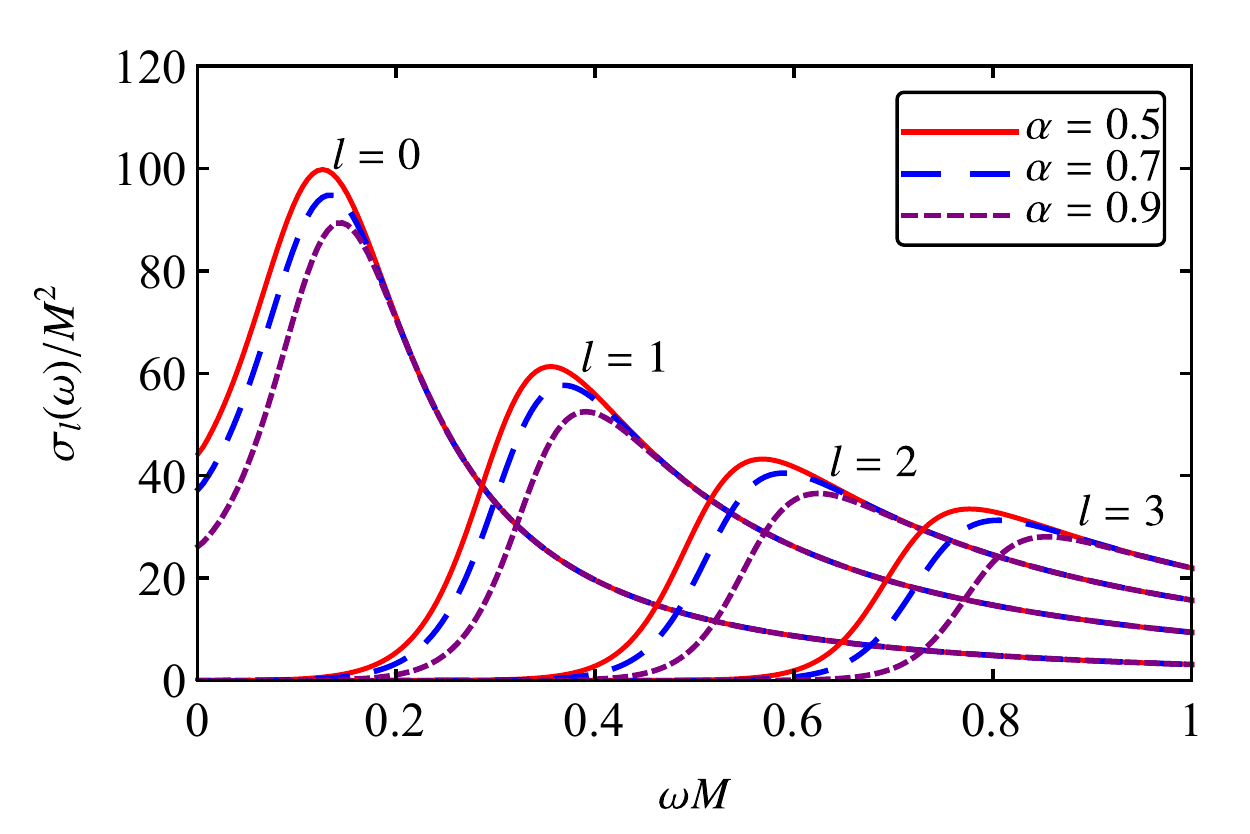}
    \caption{The partial ACS of the ABG RBH for different values of normalized charge as a function of $\omega$.}
    \label{PACSQ0}
\end{centering}
\end{figure}

\begin{figure}[!htbp]
\begin{centering}
    \includegraphics[width=\columnwidth]{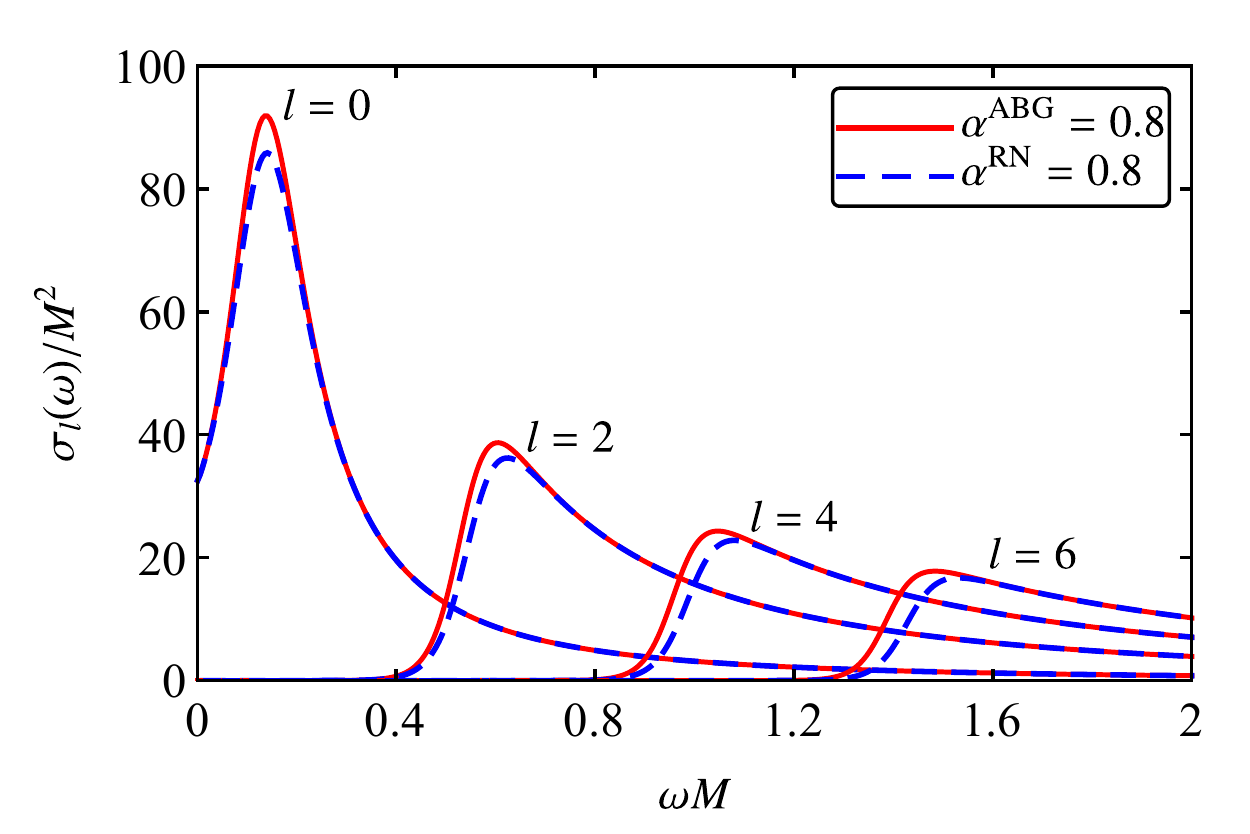}
    \caption{The partial ACSs of ABG and RN BHs for different values of normalized charge as a function of $\omega$.}
    \label{PACSABGRNQ0}
\end{centering}
\end{figure}

In Fig.~\ref{PACSQ0} we show the partial ACS of ABG RBHs for different choices of $\alpha$ as a function of $\omega$. As we can see, the partial-waves modes present a peak, which decreases as we increase $\alpha$, and vanish in the limit $\omega M \rightarrow \infty$. Moreover, for a fixed value of $\alpha$, the maximum of the partial ACS of the ABG RBH is bigger than in the corresponding RN BH case, as it is shown in Fig.~\ref{PACSABGRNQ0}.

\begin{figure}[!htbp]
\begin{centering}
    \includegraphics[width=\columnwidth]{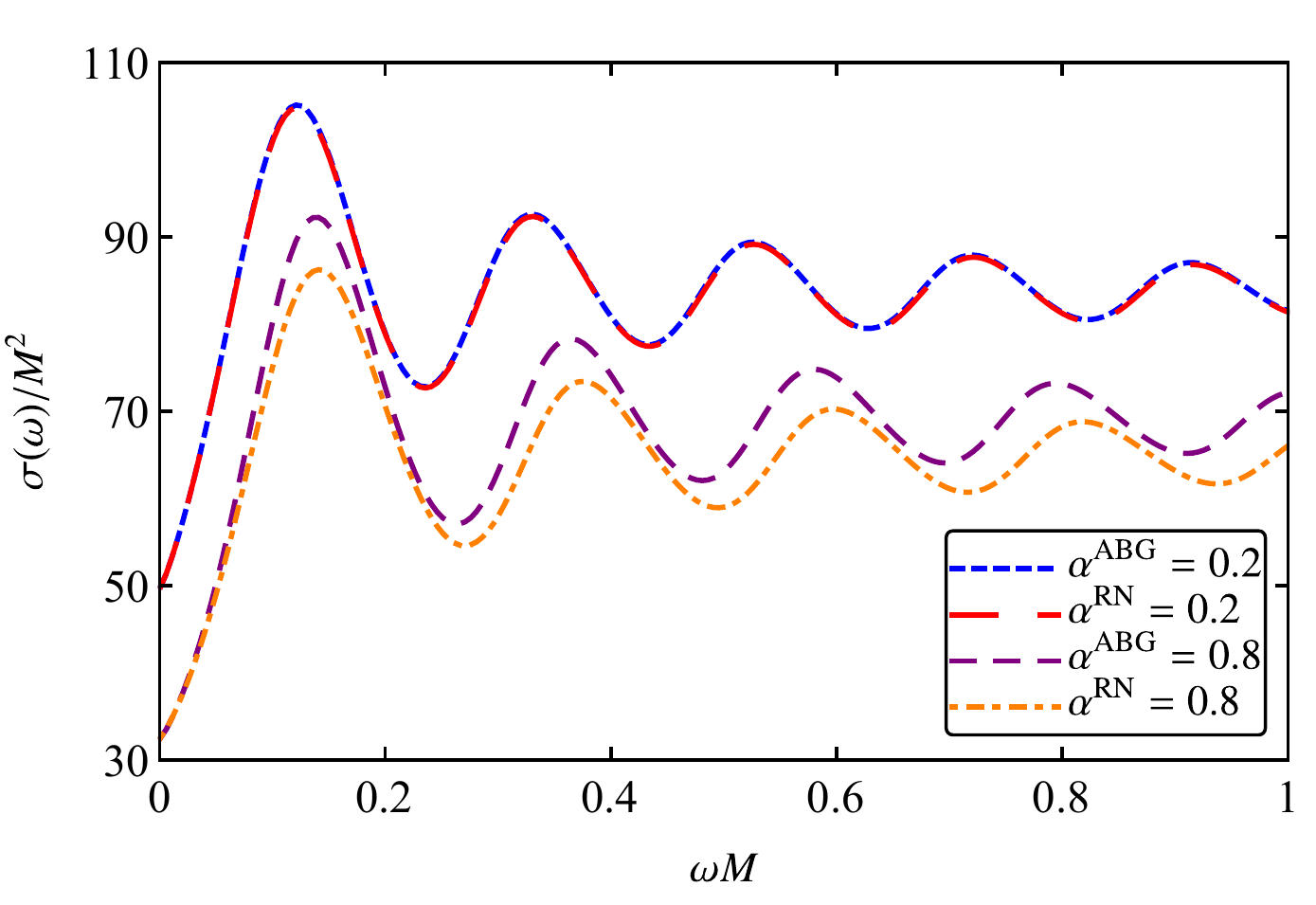}
    \caption{The total ACSs of ABG and RN BHs for two selected values of the normalized charge, as a function of $\omega$.}
    \label{TACSABGRNQ0}
\end{centering}
\end{figure}

In Fig.~\ref{TACSABGRNQ0} we present a comparison of the total ACS of ABG RBHs with the corresponding RN BHs with the same values of $\alpha$, as a function of $\omega$. For small values of $\alpha$, the total ACS of both BH solutions can be very similar along the whole frequency range. Nevertheless, as we increase $\alpha$, we note that the total ACS of the ABG RBH is typically larger than the corresponding RN case. 
The ACSs of ABG and RN BHs, for the same choice of $\alpha$, have very similar low-frequency values, in accordance with the fact that that the areas of ABG and RN BHs with the same $\alpha$ are very similar (cf. Sec.~\ref{subsec:low-frequency} and, in particular, Fig.~\ref{bharea}).

\subsection{Can RBHs mimic standard BHs?}
\begin{figure}[!htbp]
\begin{centering}
    \includegraphics[width=\columnwidth]{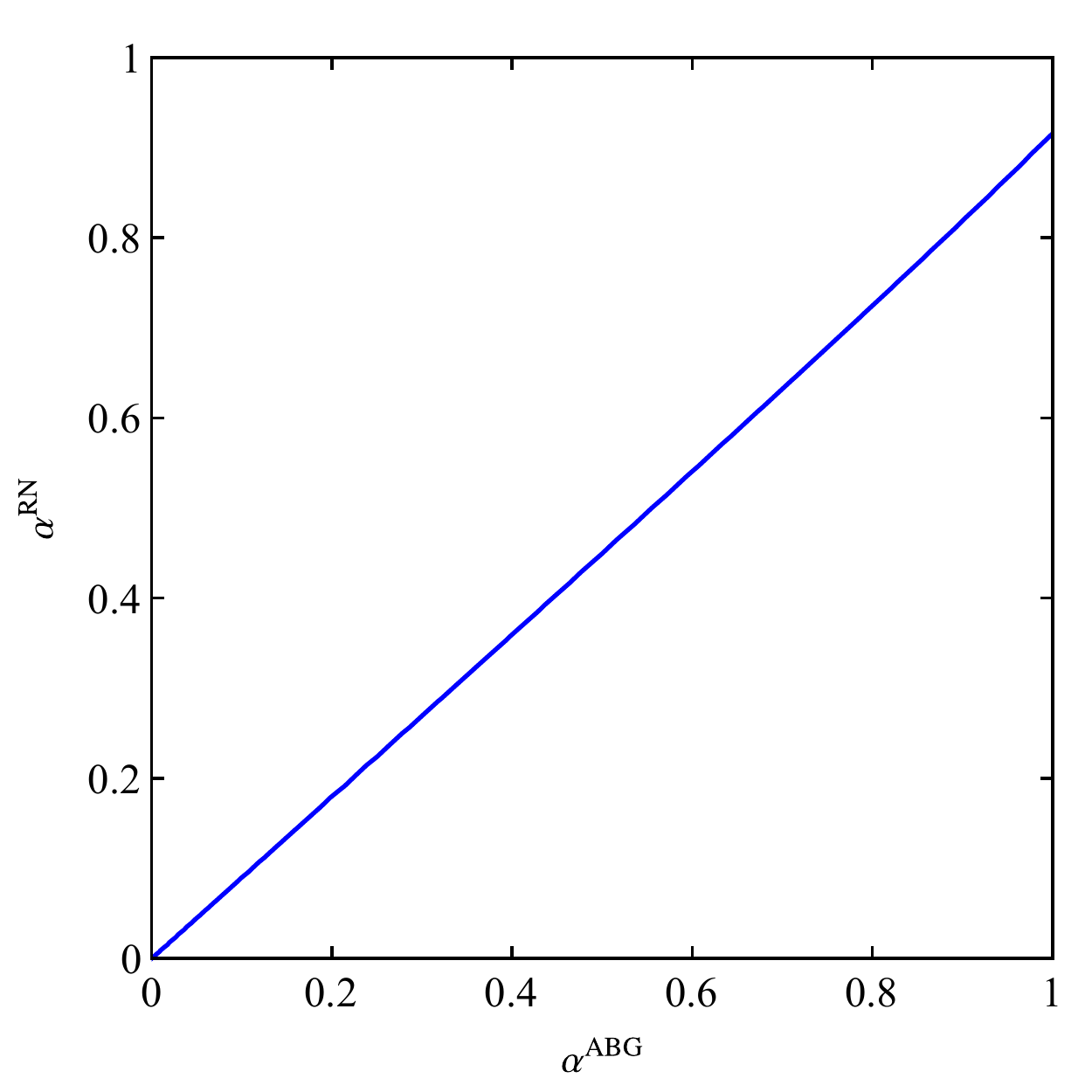}
    \caption{The values of the normalized charges for which the GCSs of ABG and RN BHs are equal.}
    \label{GCSmimic}
\end{centering}
\end{figure}

We have seen that for low values of the normalized charge, the results for the ACS of ABG BHs are similar to those of RN BHs, in the whole frequency range. This similarity opens up the possibility of RBHs mimic standard BHs solutions, when one considers the absorption results by charged BHs. We can thus search for certain values for the pair $(\alpha^{\rm{ABG}},\alpha^{\rm{RN}})$ to find situations in which the results for the ACSs are similar in the whole frequency range. A good starting point are the values of $(\alpha^{\rm{ABG}},\alpha^{\rm{RN}})$ for which the GCSs coincide. 
From Fig.~\ref{GCSmimic} we notice that the equality between the GCSs of ABG and RN BHs can be found up to $(\alpha^{\rm{ABG}},\alpha^{\rm{RN}}) = (1,0.9161)$.
We then compute the ACSs for such values, in the whole frequency regime. 

\begin{figure*}[!htbp]
\begin{centering}
    \includegraphics[width=\columnwidth]{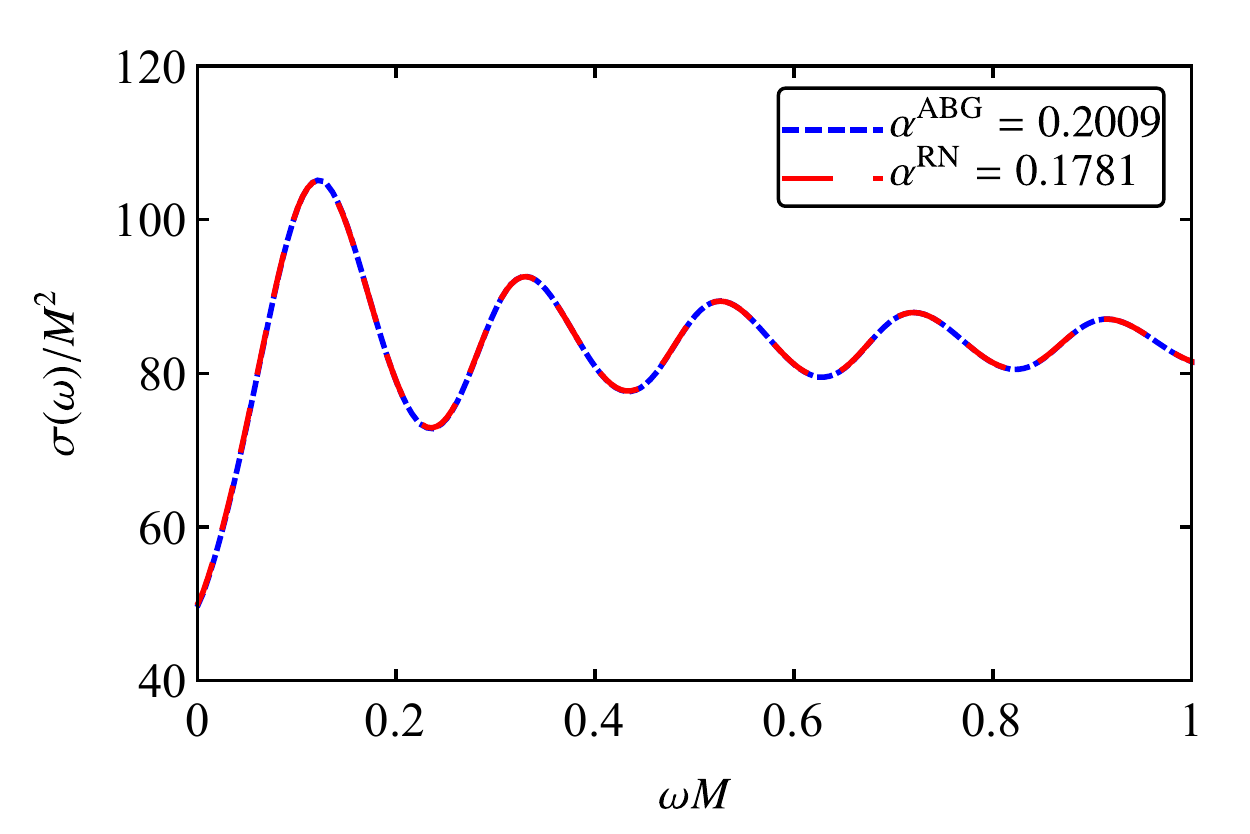}
    \includegraphics[width=\columnwidth]{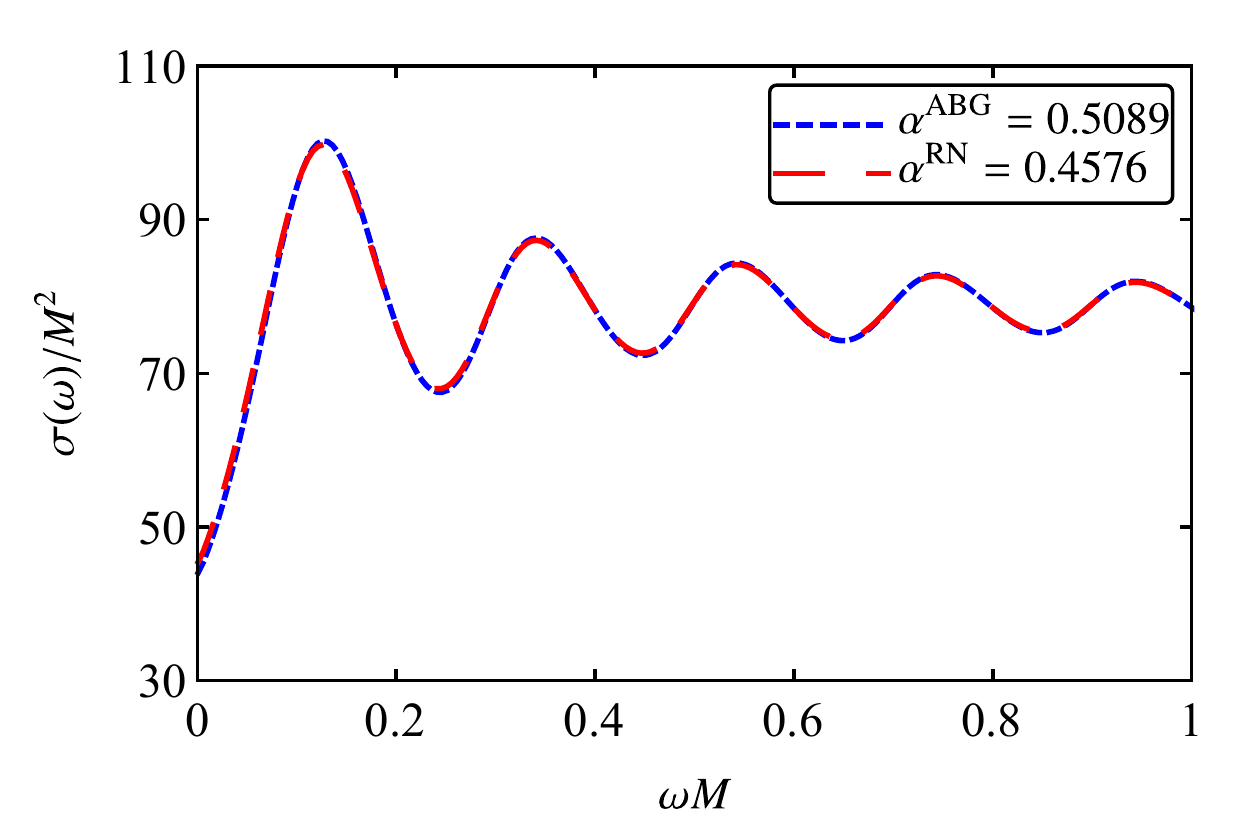}
    \includegraphics[width=\columnwidth]{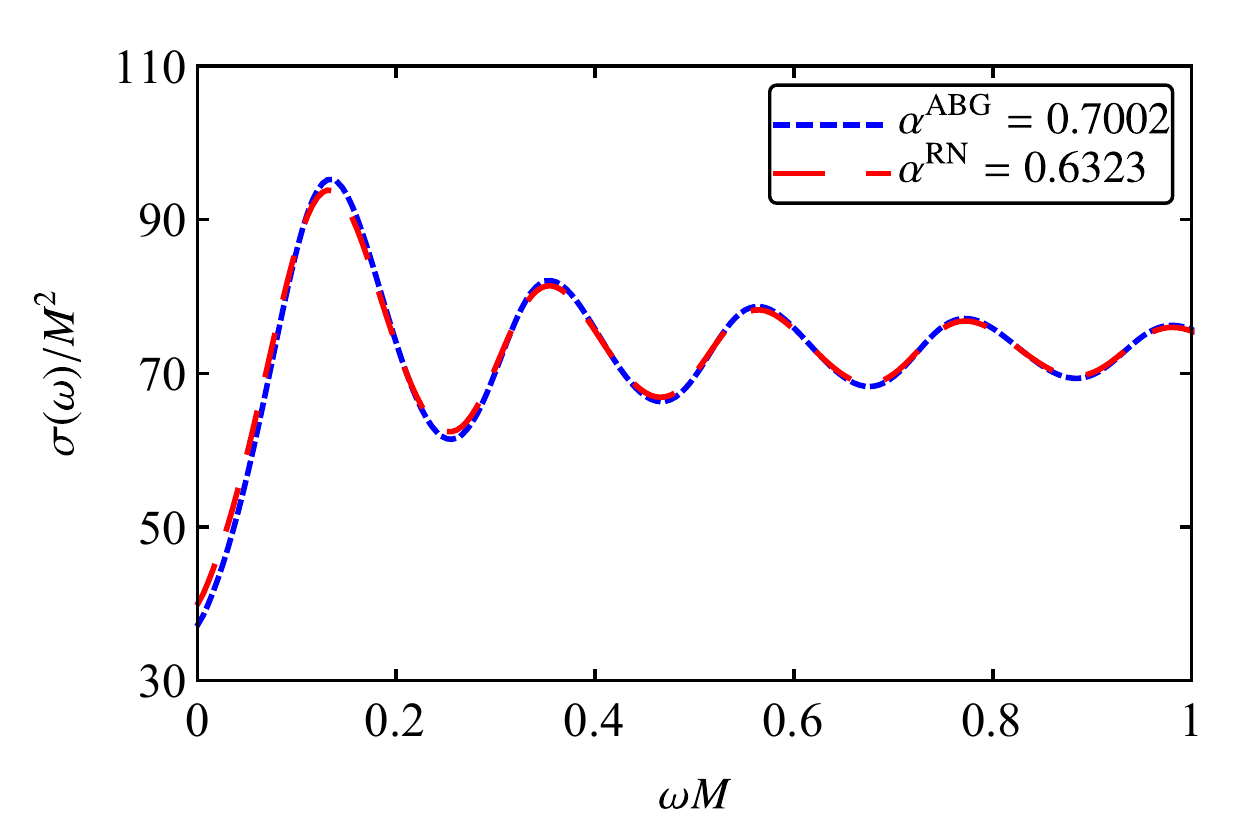}
    \includegraphics[width=\columnwidth]{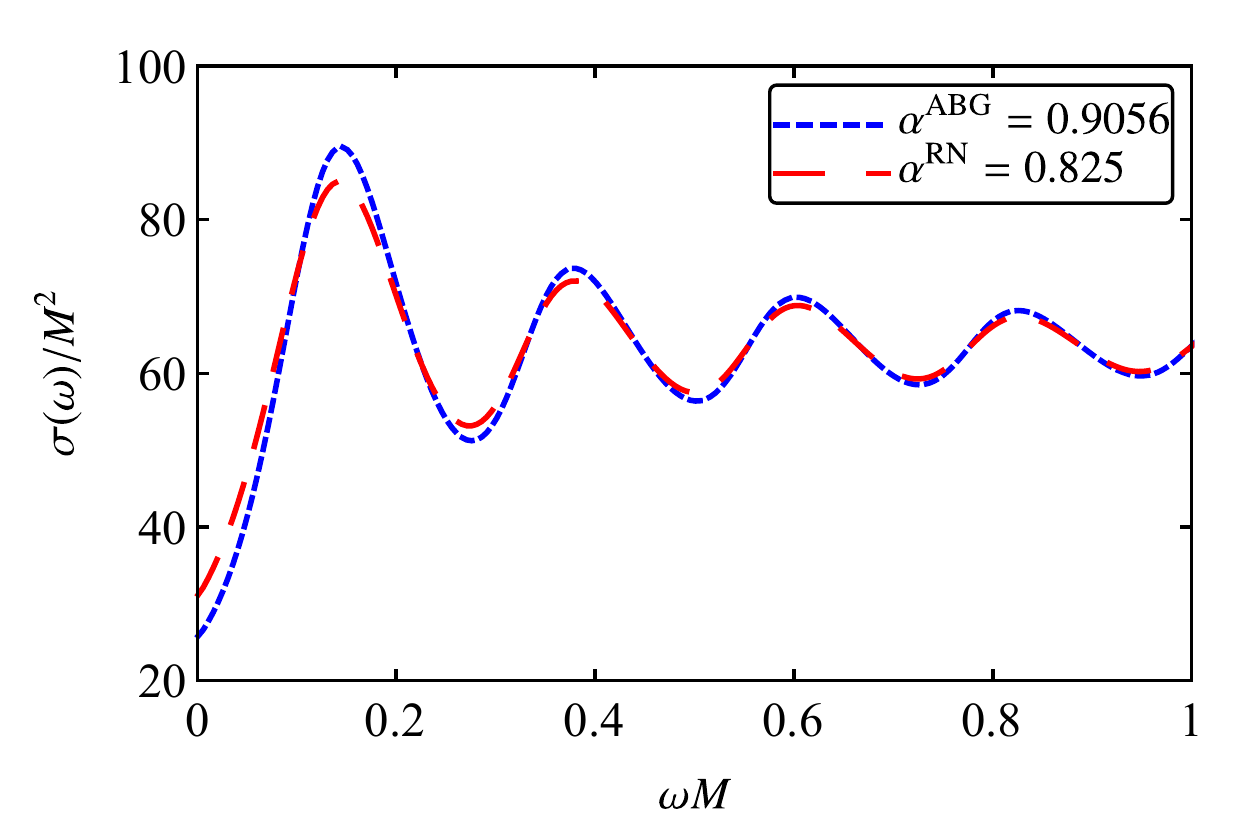}
    \caption{The total ACSs for some pairs $(\alpha^{\rm{ABG}},\alpha^{\rm{RN}})$, as a function of $\omega$.}
    \label{SACSABGRN}
\end{centering}
\end{figure*}

In Fig.~\ref{SACSABGRN} we exhibit the total ACSs for some pairs $(\alpha^{\rm{ABG}},\alpha^{\rm{RN}})$, for which the GCSs are the same. For low to moderate values of the normalized charge  $\alpha$, we observe that the total ACS of the ABG and RN BHs can be very similar for arbitrary values of the wave frequency. However, for higher values of $\alpha$, we see that the ACSs start to differ, specially in the low-frequency regime.

\section{Final Remarks}\label{sec:remarks}
We studied the propagation of a massless and chargeless test scalar field in the background of ABG RBHs, focusing in the absorption process. We computed the ACS numerically and compared our results with limiting cases, showing that they are in excellent agreement.

The metric function and the electric field of ABG RBH tend to the RN BH case in the far-field limit, but close to the event horizon they differ significantly. In particular, the quantities $f^{\rm{ABG}}(r)$ and $E^{\rm{ABG}}(r)$ are finite at the origin $(r = 0)$, whereas $f^{\rm{RN}}(r)$ and $E^{\rm{RN}}(r)$ diverge in this limit. We also note that the magnitude of the event horizon radius, $r_{+}$, of both BH solutions is very similar, for the same values of the normalized charge $\alpha$. The functions $f^{\rm{ABG}}(r)$ and $f^{\rm{RN}}(r)$ may be equal at distinct values of $r/M$.

The GCS for null geodesics of ABG RBH is typically larger than the RN one, as well as $r_{c}$ and $b_{c}$. In the chargeless case $(\alpha = 0)$, the GCSs of both ABG and RN BHs are equal to the Schwarzschild result. We obtained that the GCSs of ABG and RN BHs may also be equal for nonvanishing values of the normalized charges, and such an equality can be found for $\alpha^{\rm{RN}} \lesssim 0.9161$.

The effective potential related to the propagation of massless scalar fields in the background of the ABG RBH presents a peak which increases as we increase $l$ or $\alpha$. Generically, we note that the total ACS of the ABG RBH, in the mid-to-high frequency limit, oscillates around the corresponding GCS. Besides that, the sinc approximation provides excellent results for the ABG RBH ACS in this regime. We also note that the ABG RBH total ACS diminishes as we increase $\alpha$. This is in accordance with the fact that $V^{\rm{ABG}}_{\rm{eff}}(r)$ increases as we increase $\alpha$, what means that massless scalar waves are subject to higher potential barriers as we consider higher values of $\alpha$. Moreover, the ABG RBH total ACS is typically larger than the RN one, for the same choice of $\alpha$.

It was shown in Ref.~\cite{MC2014} that the ACSs of the Bardeen RBHs can present an oscillatory behavior similar to those of the RN BHs, in the mid-to-high-frequency regime. Here, we have shown that for for small-to-moderate values of the normalized charge, the results for the ACSs of the ABG RBH are very similar to those of the RN BH in the whole frequency range. Hence, from the perspective of the absorption of a nonmassive chargeless test scalar field by a low-charge BH, we  may not necessarily distinguish an ABG RBH from a RN BH. 

\begin{acknowledgments}
The authors would like to thank C. L. Benone and C. F. B. Macedo for useful discussions. We are grateful to Conselho Nacional de Desenvolvimento Cient\'ifico e Tecnol\'ogico (CNPq) and Coordena\c{c}\~ao de Aperfei\c{c}oamento de Pessoal de N\'ivel Superior (CAPES)—Finance Code 001, from Brazil, for partial financial support. This research has also received funding from the European Union's Horizon 2020 research and innovation programme under the H2020-MSCA-RISE-2017 Grant No. FunFiCO-777740.
\end{acknowledgments}


\end{document}